\documentclass[twocolumn]{aastex631}

\received{April 5, 2022}
\accepted{May 11, 2022}

\submitjournal{ApJ}

\shorttitle{Improving Astronomical Time-series Classification via Data Augmentation with GANs}
\shortauthors{García-Jara et al.}
\graphicspath{{./}{figures/}}
\definecolor{capri}{rgb}{0.0, 0.75, 1.0}
\definecolor{blizzardblue}{rgb}{0.67, 0.9, 0.93}
\definecolor{columbiablue}{rgb}{0.61, 0.87, 1.0}

\usepackage[ruled,vlined]{algorithm2e}
\usepackage{todonotes}
\newcounter{todocounter}
\usepackage{amsmath}
\usepackage{multirow}
\usepackage{bm}
\usepackage{mathtools}
\usepackage{lipsum}
\usepackage{enumitem}
\usepackage{dcolumn}

\newcolumntype{e}[1]{D{e}{e}{#1}}
\newcolumntype{C}[1]{D{.}{.}{#1}}

\DeclarePairedDelimiter{\ceil}{\lceil}{\rceil}

\begin{document}

\title{Improving Astronomical Time-series Classification via Data Augmentation with Generative Adversarial Networks}
% \footnote{Released on April, 5th, 2022}}

\author[0000-0002-0786-7307]{Germán García-Jara}
\affiliation{Dept. of Electrical Engineering, Universidad de Chile}

\author[0000-0002-8178-8463]{Pavlos Protopapas}
\affiliation{Institute for Applied Computational Science, Harvard University}

\author[0000-0001-9164-4722]{Pablo A. Estévez}
\affiliation{Dept. of Electrical Engineering, Universidad de Chile}
\affiliation{Millennium Institute of Astrophysics, Chile}

\begin{abstract}

% Brief description of the current scenario
Due to the latest advances in technology, telescopes with significant sky coverage will produce millions of astronomical alerts per night that must be classified both rapidly and automatically. Currently, classification consists of supervised machine learning algorithms whose performance is limited by the number of existing annotations of astronomical objects and their highly imbalanced class distributions. 
% What we propose
In this work, we propose a data augmentation methodology based on Generative Adversarial Networks (GANs) to generate a variety of synthetic light curves from variable stars.
% What we can achieve
Our novel contributions, consisting of a resampling technique and an evaluation metric, can assess the quality of generative models in unbalanced datasets and identify GAN-overfitting cases that the Fréchet Inception Distance does not reveal. We applied our proposed model to two datasets taken from the Catalina and Zwicky Transient Facility surveys. The classification accuracy of variable stars is improved significantly when training with synthetic data and testing with real data with respect to the case of using only real data.

\end{abstract}


 \section{Introduction} 
\label{sec:intro}

% DL achieve the SOTA in many taks so we would like to use it for classification of astronomical time-series. 
Deep learning models have become state-of-the-art in an extensive range of tasks, such as image recognition, video analysis, and natural language processing, demonstrating their immense ability to solve complex problems and outperform existing algorithms. Based on this fact, applying deep learning models to the classification of astronomical time-series arises as an interesting approach. 

% Practices to achieve such results
Models have progressively increased their number of parameters to achieve such results, from thousands to millions. Unfortunately, architectures with such a large number of parameters are vulnerable to overfitting. Overfitting occurs when models memorize the data available in the training set rather than learning meaningful characteristics from the data so that the model can generalize and perform well when testing on new and unseen data. To avoid overfitting, models that achieve state-of-the-art results in different tasks are trained with annotated datasets that have been extensively processed and filtered, and that consist of a large number of samples for each class, thus preventing overfitting. 
% Overfitting can be also caused by oversampling \citep{dmimbchawla2009}

% Real-world datasets (Astronomical)
However, real-world problems present different scenarios in regard to data. For example, not only there is a small number of annotations in astronomical time-series datasets, but the annotations have also highly imbalanced class distributions. While small datasets already hinder learning by making algorithms fail at generalizing characteristics of the data, imbalanced distributions only accentuate this issue (\citealt{caruana2000}; \citealt{hegarcia2009}). These two characteristics, in addition to the irregularly time-spaced nature of astronomical observations, are a considerable difficulty for machine learning algorithms and make the classification problem a unique challenge.

% What has it been done to solve it 
To overcome these problems, data augmentation techniques are frequently applied to transform small imbalanced datasets into large and balanced datasets. Most of these techniques, although widely applied in the domain of images, cannot be directly applied in the time domain due to its dissimilar properties. Consequently, augmentation techniques in the time domain remain a challenge and deserve more attention from the community \citep{timeDAforDL2020}.

% Talk about traditional techniques and why they fail 
Traditional augmentation techniques in the time domain, such as jittering, window warping, and slicing, assume that these transformations exist naturally in the data and that the augmented samples will be valid time-series with similar properties to the existing ones. Moreover, appropriate augmentation techniques are specific to the dataset \citep{empiricaltimeDA2021} and the task \citep{timeDAforDL2020}. An example of a dataset-specific technique could be jittering, where additive Gaussian noise is often used in sensor datasets. Yet this method cannot model the heteroscedastic nature of astronomical data. On the task-specific side, we could mention slicing or warping transformations that heavily discard or modify the context of the time-series, potentially altering the original samples' class information.

% Why a generative model, and why GANs
A generative model for data augmentation allows avoiding assumptions about existing transformations in the data. Since we will use the generated samples for classification, the generative model should learn the class conditional distribution of the data. Therefore, the model can learn how to generate new realistic samples directly from the data and preserve the class information simultaneously.

% Why GANs
Because of their ability to model complex real-world data and the wide success they have achieved across a variety of domains \citep{gansimbalanceCV2021}, Generative Adversarial Networks (GANs; \citealt{gans2014}) are the generative models of our choice.

While previous works have explored GAN-based data augmentation methods for classification, most have focused on the image domain (\citealt{emotion2018}; \citealt{liverfrid2018}; \citealt{xrays2018}; \citealt{weather2020}) and only a few in the time domain (\citealt{tcgan2018}; \citealt{snore2019}). Furthermore, \citet{tcgan2018} is the only work that addresses astronomical time-series generation.

To the best of our knowledge, \textbf{none of the existing approaches is suitable for our use-case}: dealing with irregularly-spaced data, allowing for both multi-class and physical parameter conditional generation, and focusing on the downstream task of classification.  In addition, the literature lacks a GAN evaluation metric to select appropriate models for classification tasks. 

In this work, we propose a GAN-based data augmentation methodology for time-series to improve the classification accuracy on two astronomical datasets taken from the Catalina and Zwicky Transient Facility surveys. The main contributions are: 
\begin{enumerate}[noitemsep,topsep=6pt,label=(\alph*)]
    \item 
    Proposing a GAN model capable of performing conditional generation based on class and physical parameters, suitable for irregularly-spaced time-series. 
    \item \label{contr:overfitting} 
    Revealing the incapability of the standard GAN evaluation metric (FID) to assess overfitting and proposing a novel evaluation metric that overcomes this issue to select an adequate generative model.
    \item 
    Proposing a resampling technique to delay the occurrence of overfitting. 
    \item 
    Designing two new data augmentation techniques for time-series that produce plausible time-series preserving the properties of the original ones.  
\end{enumerate}

The remainder of the paper is structured as follows: Section \ref{sec:backg} presents a theoretical background of the work. In Section \ref{sec:data} the utilized datasets and their pre-processing are explained. Section \ref{sec:meth} explains the proposed methodology. Section \ref{sec:results} presents the obtained results, which are discussed in  Section \ref{sec:discussion}, stating its strengths and weaknesses. Finally, Section \ref{sec:conclusions} presents the main conclusions of this work and future steps.

\section{Background}
\label{sec:backg}

\subsection{Imbalanced datasets}

% Describe what imbalance is 
Let $\mathcal{D} = \{x_i, y_i\}_{i=1}^{n}$ be a dataset where $x_i$ is a real example and $y_i \in Y=\{1,2,...,c\}$ a class label associated to $x_i$. $\mathcal{D}$ is said to be imbalanced if the distribution of $Y$ differs significantly from the discrete uniform distribution $\mathcal{U}\{1,c\}$. Therefore, imbalanced datasets are composed of one or more classes (majority classes) that severely outrepresent other existing classes (minority classes) \citep{hegarcia2009}.

Given an imbalanced dataset $\mathcal{D}$, we can apply sampling techniques to transform its class distribution into a uniform. The result of this transformation is a modified version of the original dataset, its balanced counterpart $\mathcal{D}^u$.

% Typically, the use of sampling methods in imbalanced
% learning applications consists of the modification of an
% imbalanced data set by some mechanisms in order to
% provide a balanced distribution. Studies have shown that
% for several base classifiers, a balanced data set provides
% improved overall classification performance compared to
% an imbalanced data set 
% Why is it a problem 
% 1)The problem of imbalanced domains is associated with a mismatch between the importance assigned by the user to some predictions (1) and the representativeness of the values involved in these predictions on the available training

% 2) He et al about failing to capture the distribution 

\subsection{Generative Adversarial Networks}

% General definition without formulas 
The GAN framework consists of a game between two networks. Given an input dataset of real samples $x_r \sim P_r$, the \emph{generator} network ($G$) aims to implicitly approximate the data distribution  $P_r$ by performing a mapping between a source of noise and the real sample space. The result of this mapping are fake samples $x_g \sim P_g$ that attempt to resemble the real ones. In contrast, the \emph{discriminator} network ($D$) tries to distinguish between $x_r$ and fake samples generated by $G$. 

During the training process, the two networks compete against each other without having control of the opponent's parameters. On the one hand, $G$ is trained to generate samples that resemble the real ones,  while on the other hand $D$ is trained to predict whether a given sample comes from the input dataset or was generated by $G$. At the end of the training, $G$ will generate samples similar to the ones in the input dataset, and the $D$ will be unable to tell apart generated from real samples.

% GANs in computer vision, wihout many details
Since the creation of GANs, they have revolutionized the field of generative modeling, showing novel results especially in the domain of images. As a broad overview of the evolution process, we could mention conditional-generation models (\citealt{cgans2014}; \citealt{acgans2017}), models that stabilize the erratic behavior of the original GANs (\citealt{wgans2017}; \citealt{wgansgp2017}; \citealt{spectralnorm2018}), and models that generate samples with an impressively high quality and resolution (\citealt{progan2018}; \citealt{biggan2018}; \citealt{stylegan2019}; \citealt{stylegan22020}) among many other models and applications. An extensive description of GAN models in computer vision is provided in \citet{reviewGANsCV2021}.

% GANs in time-series
GANs have also been applied to the time-series domain, with significant improvements in recent years. The first model capable of generating continuous sequential data was proposed by \citet{continuosrnngan2016} adding recurrent neural networks to the GAN’s generator and discriminator to handle the time-series’ temporal evolution. This work was followed by \citet{rcgan2017} who added label-conditional generation and a focus on downstream medical tasks. More recently, \citet{timeseriesgan2019} introduced a jointly trained embedding network that combines the unsupervised GAN framework with a supervised autoregressive model to capture the time-series’ conditional temporal dynamics.
Lately, \citet{sigwgan2020} proposed a GAN framework to deal with long time-series data based on an approximation of the Wasserstein distance using the signature feature space, avoiding the usage of costly discriminators and claiming to achieve state-of-the-art results in measures of similarity and predictive ability.

% Giorgia
The most related work corresponds to the T-CGAN \citep{tcgan2018}, which proposes a method to generate irregularly-spaced time-series. Still, it does not include conditional generation with physical parameters of interest, it does not perform multi-class generation, and similarly to the works mentioned above, it does not tackle the problem of model selection for a downstream task.

\subsubsection{Wasserstein GAN}
\label{subsubsec:WGAN}

The Wasserstein-GAN (WGAN, \citealt{wgans2017}) is one of the GAN models that is widely used and well known for its training stability. This GAN leverages an approximation of the Wasserstein-1 distance to measure the dissimilarity between $P_r$ and $P_g$. An upgraded version of this model is the WGAN with Gradient Penalty (WGAN-GP, \citealt{wgansgp2017}), which adds a regularization term to the original WGAN loss to satisfy the Lipschitz condition on $D$. The WGAN-GP objectives that are minimized during the training process are: 

\begin{eqnarray}
\label{eq:lg}
L_G &=& \mathop{\mathbb{E}}\limits_{x_r \sim P_r}[D(x_r)] - \mathop{\mathbb{E}}\limits_{x_g \sim P_g}[D(x_g)] \\
\label{eq:ld}
L_D &=& -L_G + \lambda \mathop{\mathbb{E}}\limits_{\hat{x} \sim P_{\hat{x}}}[(\lVert\nabla_{\hat{x}}D(\hat{x})\rVert_2 - 1 )^2] 
\end{eqnarray}

\noindent
where $P_{\hat{x}}$ is the distribution implicitly defined by sampling uniformly along linear paths between points sampled from $P_r$ and $P_g$, and $\lambda$ is the penalty coefficient that controls the strength of the gradient regularization. 

\subsection{Data Augmentation}

% Define DA
Data augmentation refers to a set of techniques applied to a dataset used to create new samples that are slightly different from the existing ones to increase the number of samples in the dataset. It is frequently used to prevent overfitting, and it helps improve the performance of machine learning models for various applications \citep{timeDAforDL2020}. 
Classic examples of this in the field of images are rotations, translations, crops, and flips, among others. 

% DA in the time-domain
In the time domain, data augmentation techniques are less standardized. Traditional techniques in time-series correspond to non-parametric transformations such as jittering, scaling, window-slicing, and window-warping \citep{daclfcnn2016}. Parametric techniques can also be applied in data augmentation, such as the parametric model-based augmentation for transient phenomena proposed in \citet{pimentel2022deep}.

\subsection{Overfitting in GANs}
\label{subsec:of}
 As described in \citet{ada2020}, overfitting in GANs occurs when training on small datasets. The less data there is, the earlier the discriminator becomes too confident in separating real from generated samples, which impedes the progress of $G$ and eventually deteriorates the quality of the generated samples.

Even though \citet{ada2020} proposed Adaptive Discriminator Augmentation (ADA) as a technique to deal with overfitting in GANs, this technique requires the application of differentiable transformations to augment the training data. Since our goal is to provide a GAN-based data augmentation method motivated by the limited augmentation methods for time-series, we intentionally do not include any augmentation method (apart from oversampling) in the GAN-training process, hence we do not consider using ADA. 

\subsection{Evaluation of GANs}
\label{subsec:evalgan}

% Why is evaluation necessary ?
Even though the losses described in Equations \ref{eq:lg}, \ref{eq:ld} successfully describe the adversarial problem and quantify the distance between $P_r$ and $P_g$, their high variance makes them unsuitable for using them as a stopping criterion. Even if they did not suffer from this issue, metrics based on $D$ are specific to their corresponding $G$, and cannot generalize properties about the generated dataset. Consequently, the framework requires additional evaluation metrics to assess the quality of the generated samples and select the definitive generator for the downstream task.

Evaluation of generative models requires a notion of the distance between $P_r$ and $P_g$. Defining such a measure for high dimensional distributions is a challenging task and remains an open problem \citep{dc2020}. 

An intuitive way of comparing these distributions is as follows: if a generative model can successfully capture $P_r$ with $P_g$, the performance on any downstream task should be similar when our data comes from any of the two distributions. Setting the downstream task to classification leads to using classification metrics for evaluation.

\subsubsection{Classification metrics}

Considering that the ultimate purpose of this work is to improve the classification of real astronomical objects, we naturally adopt the classification accuracy metric first proposed in \citet{lr2017} and later used in \citet{rcgan2017}, \citet{covariategan2018}, \citet{howgood2018} and \citet{cas2019}. For clarity, we choose to preserve the names in \citet{rcgan2017}: Train on Synthetic Test on Real (TSTR) and Train on Real Test on Real (TRTR). These two scores are computed by training a classifier on synthetic (generated) data or real data and then evaluating its classification accuracy on real data. 

\subsubsection{Feature-based metrics}
\label{subsubsec:featmetrics}

Based on the difficulty of finding meaningful metrics in the input space, quantifying the distance between the distributions $P_r$ and $P_g$ often involves mapping samples $x \in \{x_r, x_g\}$ into a feature space with a transformation $x \mapsto \phi(x)$, where $\phi$ is an intermediate representation of a pre-trained classifier (\citealt{is2016}; \citealt{ttur2017}; \citealt{pr2018}; \citealt{impr2019}; \citealt{dc2020}). The classifier is generally a convolutional neural network (CNN) such as the Inception-v3 \citep{inception2016}, a widely used architecture in computer vision. 

Since the dimensionality of $\phi$ is often lower than that of $x$, the distributions of the feature space are often called \textit{manifolds}. We will informally understand these manifolds as connected regions with a relatively simple structure embedded in a more complex space. 

When evaluating generative models, two desired characteristics are \textbf{fidelity} and \textbf{diversity}. The former describes how real the generated samples look in comparison to the real ones, while the latter measures how much of $P_r$ the model can cover with $P_g$.

  \paragraph{The Fréchet Inception Distance (FID)} This metric proposed by \citep{ttur2017} consists of a Wasserstein-2 distance between $\Phi_r$ and $\Phi_g$, the distributions of $\phi_{r}$ and $\phi_{g}$ respectively.
  
Under the assumption that both distributions are multivariate Gaussians, their  mean $\mu$ and covariance $\Sigma$ are estimated to obtain a closed-form of the distance: 

\begin{equation}
    \label{eq:fid}
    FID = {\underbrace{||\mu_r - \mu_g||}_\text{(a)}}^2 + \text{Tr} (\underbrace{\Sigma_r + \Sigma_g - 2 (\Sigma_r \Sigma_g)^{1/2}}_\text{(b)})
\end{equation}

% Briefly describes what the formula means.
While (a) can be interpreted as a measure of fidelity that indicates the average distance between the two distributions, (b) can be interpreted as a measure of diversity that compares the variability of the two distributions.

A particularly relevant limitation of FID in the presence of highly imbalanced distributions is that computing the last term in (b) requires full-rank $\Sigma$ matrices, which makes the calculation of a per-class FID unfeasible if the minority classes contain fewer samples than the dimensionality of $\Phi$. Furthermore, even if we had enough samples to compute it, a per-class score would be unreliable for the minority classes since FID is known to suffer from high bias for small sample sizes \citep{demystifying2018}.

\paragraph{Precision and Recall} \citet{pr2018} proposed separating fidelity and diversity into two relative-density-based metrics: \textbf{precision} and \textbf{recall}. These two metrics improve upon FID by identifying cases of mode dropping or mode inventing in the generated distribution, in the pathological case where different models achieve similar FID values by privileging either one of the two terms in Equation \ref{eq:fid}.

\paragraph{Improved Precision and Recall} Motivated by the failure at identifying models with poor variability, \citet{impr2019} proposed improved precision and recall metrics (\textbf{P\&R}). These metrics are computed by estimating the manifolds $\Phi \in \{\Phi_r, \Phi_g\}$ according to:

\begin{eqnarray}
\label{eq:manifold}
\hat{\Phi} = \bigcup_{\phi \in \mathbf{\Phi}} B(\phi,NND_k(\phi))
\end{eqnarray}

\noindent 
where $\mathbf{\Phi} \in \{\mathbf{\Phi}_r, \mathbf{\Phi}_g\}$ is a collection of feature samples $\phi \in \{\phi_r, \phi_g\}$, the ball $B(x,r)$ is the solid sphere around $x$ with radius $r$, and $NND_k(\phi)$ is the distance from $\phi$ to its $k$-th nearest neighbor within the corresponding manifold. In the presence of outliers, the KNN approach results in an over-estimation of the manifolds due to the large distances between samples.

\paragraph{Density and Coverage} \citet{dc2020} proposed \textbf{density} and \textbf{coverage} (\textbf{D\&C}) motivated by the vulnerability of \textbf{P\&R} to outliers. While \textbf{P} measures fidelity depending on the binary decision of whether a feature sample $\phi_g$ belongs to the real manifold $\Phi_r$, \textbf{D} considers the amount of balls $B(\phi_r,NND_k(\phi_r))$ within each $\phi_g$ is contained, adding robustness to real distributions with outliers. On the other hand, \textbf{C} measures diversity based on the real manifold estimate instead of the generated one, in contrast to \textbf{R}. 

In our practical case, we found that \textbf{P} and \textbf{C} saturate quickly, not providing meaningful information.  Since these metrics directly depend on the real manifold estimates, we hypothesize that this behavior can be caused by the sparsity of $\Phi_r$ in the minority classes, leading to the same over-estimation issue as outliers. Consequently, we decide to use \textbf{D} and \textbf{R} as our fidelity and diversity metrics.

Let $B_r^k$ be the abbreviation of $B(\phi_{r},NND_{k}(\phi_{r}))$, and $\hat{\Phi}_g$ the approximation of the generated manifold described in Equation \ref{eq:manifold},  we compute the \textbf{D\&R} metrics according to: 

\begin{eqnarray}
\mathbf{D}_{(\mathbf{\Phi}_{r},  \mathbf{\Phi}_{g})}&=& \displaystyle \frac{1}{k|\mathbf{\Phi}_{g}|} \sum_{\phi_{g} \in \mathbf{\Phi}_{g}}  \sum_{\phi_{r} \in \mathbf{\Phi}_{r}}\mathbf{1}_{B_r^k}(\phi_g) \label{eq:density} \\
\mathbf{R}_{(\mathbf{\Phi}_{r},  \mathbf{\Phi}_{g})}&=& \displaystyle \frac{1}{|\mathbf{\Phi}_{r}|} \sum_{\phi_{r} \in \mathbf{\Phi}_{r}}\mathbf{1}_{\hat{\Phi}_g}(\phi_r) \label{eq:recall} \end{eqnarray}

\noindent
where $\mathbf{1}_{A}(x)$ is the indicator function defined as: 

\begin{equation}
    \mathbf{1}_{A}(x)={\begin{cases}1~&{\text{ if }}~x\in A~,\\0~&{\text{ if }}~x\notin A~.\end{cases}}
\end{equation}

\section{Data} 
\label{sec:data}

\subsection{Datasets}
\label{subsec:datasets}

Because of the recognizable shapes of their light curves when visualized in phase space, we focus on periodic variable stars. However, the framework could be effortlessly extended to other stars of interest if needed. We perform and validate our experiments on data captured by two time-domain astronomical surveys.

\paragraph{The Catalina Surveys Data Release-1}

This catalog described in \citet{catalina2014}, captured with the $8.2 \ deg^{2}$ field-of-view camera mounted on the CSS 27-inch Schmidt telescope, provides $\sim$61,000 light curves of periodic variable objects, with their corresponding periods and classes. To decrease the complexity of the multi-class problem induced by the large number of periodic classes provided, we only consider a subset of the periodic objects grouped following the mapping described in Table \ref{table:CatalinaMapping}.

\begin{table}[!htbp]
\centering
\normalsize
\caption{Adopted classes distribution for the Catalina Surveys Data Release-1. The original class acronyms as described in \citet{catalina2014} are shown in ($\cdot$).}
\label{table:CatalinaMapping}
    % \resizebox{\columnwidth}{!}{%
        \begin{tabular}{ll} 
            \hline \hline 
            {New class} & {Original class} \\ \hline
            \multirow{2}{*}{EBSD/D} & Contact eclipsing binary (EW) \\ 
            {} & {Semi-detached eclipsing binary ($\beta$ Lyrae)} \\ \hline
            \multirow{4}{*}{RRL} & {Fundamental mode RR Lyrae (RRab)} \\
            {} & {First over-tone mode RR Lyrae (RRc)} \\
            {} & {Multi-mode RR Lyrae (RRd)} \\
            {} & {Long-term modulation (Blazkho)} \\ \hline
            EBC & {Detached eclipsing binary (EA)} \\ \hline
            LPV & {Long period variables (LPV)} \\ \hline
            \multirow{2}{*}{DSCT} & {High amplitude $\delta$ Scuti (HADS)} \\ 
            {}  & {Low amplitude $\delta$ Scuti (LADS)} \\ \hline
            \multirow{2}{*}{CEP} & {Anomalous Cepheids (ACEP)} \\
            {} &  type-II Cepheids (Cep-II) \\
        \end{tabular}
    % }
\end{table}

\paragraph{The Zwicky Transient Facility}

This survey, known by its acronym ZTF \citep{zwicky2018} provides a public multiband stream of alerts captured by a $47 \  deg^{2}$ field-of-view camera mounted on the Palomar 48-inch Schmidt telescope, is capable of scanning the entire northern sky every three nights and the plane of the Milky Way every night. To enable further analysis in follow-up telescopes, the alerts are processed by alert brokers that are designed to provide a rapid and self-consistent classification.
We use the subset of periodic variable stars present in the ZTF training set created by the ALeRCE broker \citep{alerce2021}, along with their taxonomy. This training set was constructed considering sources observed by ZTF whose labels had been cross-matched from different multiple catalogs. 

Previous works (\citealt{late2021}, \citealt{carrasco2021stamp}) have already used ZTF data processed by the ALeRCE broker to train different machine learning algorithms. More details about the data processing can be found in \citet{alerce2021}.

After pre-processing both datasets following the steps detailed in Section \ref{subsec:preprocess}, we obtain the definitive versions of the datasets that will be used in our experiments, from now on referred to as the ``Catalina" and the ``ZTF" datasets. The class distributions of the pre-processed datasets are shown in Table \ref{table:ClassDistribution}.

\begin{table}[!htbp]
\centering
\normalsize
\caption{Class distributions of the pre-processed datasets.}
\label{table:ClassDistribution}
    % \resizebox{\columnwidth}{!}{%
        \begin{tabular}{ll|ll} 
            \hline \hline 
            \multicolumn{2}{c|}{\textbf{Catalina}} & \multicolumn{2}{c}{\textbf{ZTF}} \\
            Class & $N \textsuperscript{\b{o}}$ samples & Class & $N \textsuperscript{\b{o}}$ samples \\
            \hline
            EBSD/D & 28980 & EB & 31477 \\
            RRL & 7533 & RRL & 18729 \\
            EBC & 4500 & LPV & 5245 \\
            LPV & 483 & DSCT & 507\\
            DSCT & 241 & CEP & 471\\
            CEP & 182
        \end{tabular}
    % }
\end{table}

\subsection{Data pre-processing}

\label{subsec:preprocess}

To use the data described in Section \ref{subsec:datasets}, some pre-processing steps need to be applied. The pre-processing consists of four main steps: period folding, outlier filtering, time sampling, and median centering.

\subsubsection{Period folding}

Since the desired characteristic shapes of periodic light curves are only visible in the phase space, we start by folding the light curves into the period provided in both datasets. Denoting the light curve period as $T$, and the observation time as $t$, the folding operation is performed by converting $t$ into $\phi_t$ according to:
\begin{eqnarray}
\phi_T &\equiv & t \pmod{T} \label{eq:modulo}\\
\phi_t &=& \frac{\phi_T}{T}
\end{eqnarray}

\noindent
where the congruence symbol $\equiv$ in Equation \ref{eq:modulo} refers to the modulo operator with modulus $T$.
% or 
% where the congruence symbol $\equiv$ in Equation \ref{eq:modulo} refers to the remainder after dividing $t$ by $T$.

With this operation, we transform times with a variable range of values to phases with values bounded between 0 and 1. This transformation is convenient because multiple neural networks will process the phases, and having inputs with a similar range is a desirable property when training such algorithms. 

\subsubsection{Outlier filtering}
\label{subsubsec:OutFilter}

Considering that some of the light curves in the datasets can include a significant amount of noise, we filter out anomalous observations within each curve of both datasets. These anomalous observations are in general isolated observations with a magnitude that does not follow the general behavior of the magnitudes in the light curve, and including them could be detrimental to the performance of our algorithms. For the Catalina dataset, the anomalous behavior is quite particular to each light curve, and a general threshold filtering cannot be applied; therefore, a different approach is needed.

The Catalina light curves are filtered by comparing each magnitude with the local statistics of the magnitudes' neighborhood. This comparison is performed using the \textit{z-score}\footnote{The z-score is the distance of an observed value $x$ to the population mean $\mu$, measured in terms of the population standard deviation $\sigma$. It is computed by $z = \frac{x-\mu}{\sigma}$} of the magnitudes within a window that considers only a portion of the light curve. The process is performed by sliding the window through the entire light curve with a window size $w_s=20$, removing the outlier observations that satisfy $z_{score}>3$, and repeating two times per light curve since consecutive outlier observations can significantly alter the moving window's statistics and not be detected in a single pass. The results of this filtering step are shown in Figure \ref{fig:FilteringResults}b. After this step, we perform a second filtering stage by discarding the light curves that contain more than 90\% of their magnitudes out of the range delimited by the class medians and class standard deviations.

% We could say that a small window size would be too sensitive at marking outliers, and also that performing the filtering just once can fail at removing multiple outlier observations that are consecutive (multiple outliers in groups modify the local statistics so the z-score does not mark them as outliers)

On the other hand, anomalous observations in the ZTF data have been already marked with a magnitude of 100. Hence, these observations can be filtered out by a simple threshold. Following the filtering steps used by \citet{late2021}, we use $mag_{thr}=30$.

\subsubsection{Time sub-sampling}

To bring the problem to a more straightforward domain, we set the length of the light curves to a predefined value for each dataset. With this simplification, we can work with convolutional architectures rather than recurrent architectures that could hinder the GAN's training stability by violating the Lipschitz constraint, adding extra complexity to the problem.

Given a light curve with an arbitrary number of $m$ observations, we obtain the fixed-length light curves by randomly choosing $n$ from the $m$ available observations. Considering that we choose our points with no particular bias, this approach should give a reasonable approximation of the original light curve if $n$ is not too small compared to $m$.

\begin{figure}[!htbp]
\centering
\includegraphics[width=.49\textwidth]{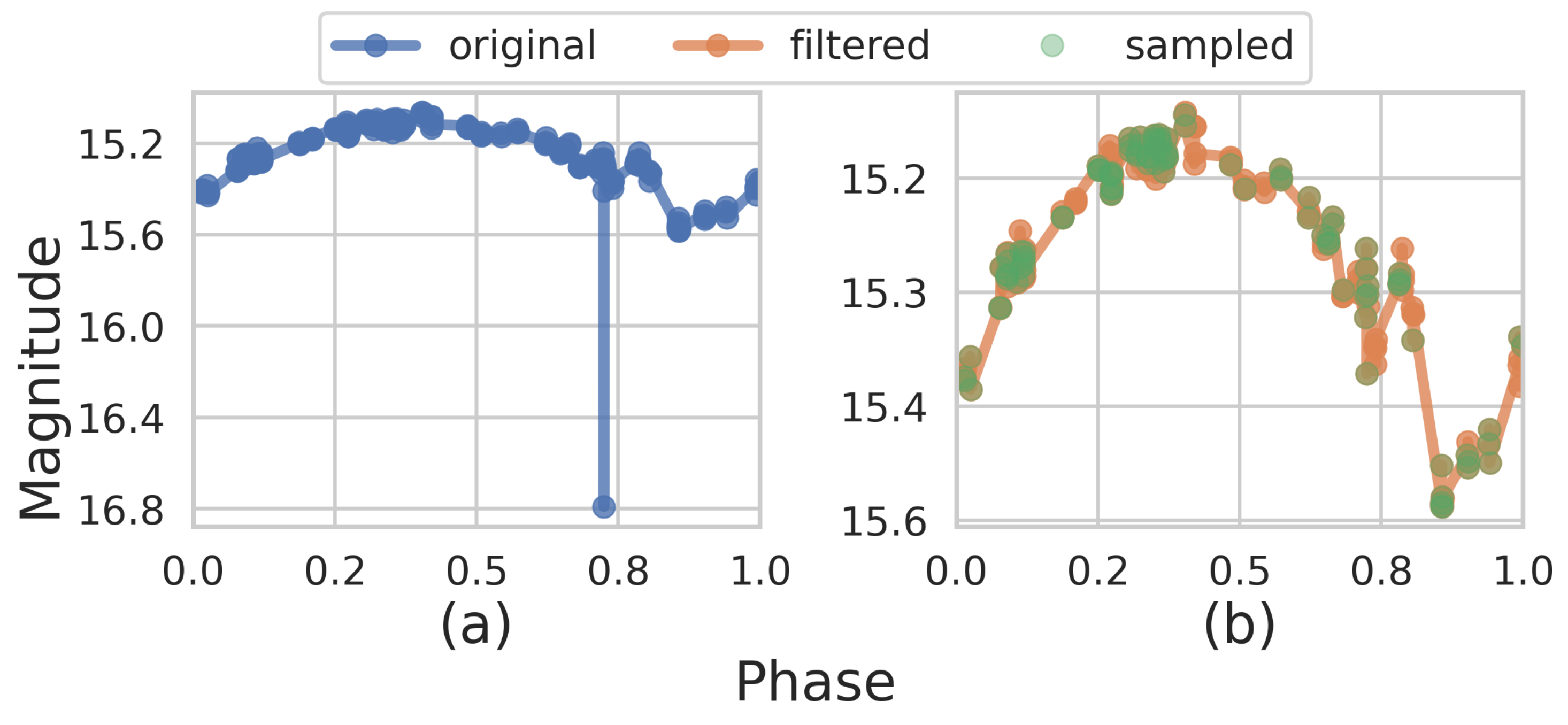}
\caption{(a) Original cepheid from the Catalina dataset. (b) Filtered and sub-sampled versions of the original cepheid.}
\label{fig:FilteringResults}
\end{figure}

Since both of our real datasets contain irregularly sampled light curves, and we perform the sub-sampling step after the period folding step,  choosing an observation implies selecting a magnitude with its corresponding observation phase. Both magnitudes and phases are part of the input of our models, as will be detailed in Section \ref{sec:meth}. Figure \ref{fig:FilteringResults}b shows an example of the time sub-sampling step.  

The light curve length is set to 100 observations for the Catalina dataset, whereas that of the ZTF dataset is 40 observations, consistent with the fact that ZTF is a relatively new sky survey with a lower number of observations per object compared to the Catalina Survey. 

After discarding the light curves that do not have the minimum length to perform this step, we end up with approximately $41k$ and $56k$ samples in the Catalina and ZTF datasets, respectively, whose class distribution is shown in Table \ref{table:ClassDistribution}.

\subsubsection{Median centering}

The last step to get the data ready for data generation is centering it around 0 so all the magnitudes have a consistent range that can be learned from the generator. This is done for each light curve by subtracting the center (median) of the magnitudes. We compute the median instead of the mean because of its robustness to outlier magnitudes.

This step is necessary because $G$ is a neural network that outputs a \textit{tanh} activation, and it can only generate values in a symmetrical range around zero. It is worth mentioning that we could center the data around any other offset, which would require to also include that offset to the output of the generator; the importance of performing this step is not the value of the offset itself, but rather the unification of all the magnitudes around a single value so our generator can model them.

\section{Methodology}
\label{sec:meth}

\subsection{General Description}
\label{subsec:graldescription}

We propose a conditional generation approach that extends the T-CGAN \citep{tcgan2018}, adding the class and amplitude of the light curves to the conditional parameters, which include the observation phases according to the original model. The details of how the conditional parameters are included into the model will be explained in Section \ref{subsec:dsdetails}.

\begin{figure}[!htbp]
\centering
\includegraphics[width=.49\textwidth]{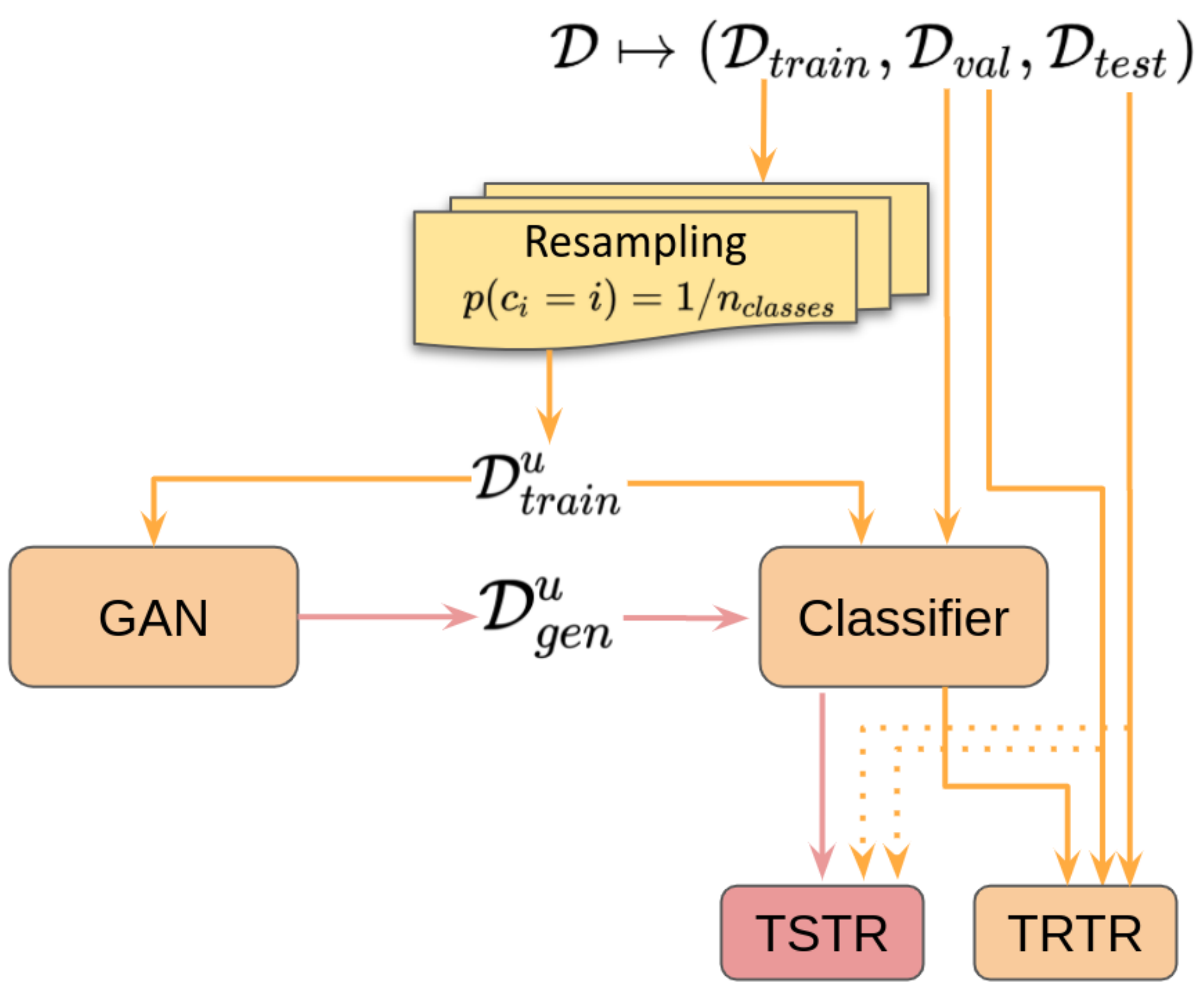}
\caption{Diagram of the methodology}
\label{fig:Flowchart.pdf}
\end{figure}

A summary of the proposed methodology, that details the partitions of datasets for the models and metrics is provided in Figure \ref{fig:Flowchart.pdf}.

We start by partitioning the pre-processed dataset $\mathcal{D}$ into $\mathcal{D}_{train}$, $\mathcal{D}_{val}$ and $\mathcal{D}_{test}$, the  \textit{train, validation}, and \textit{test} sets. Each class in $\mathcal{D}_{val}$ and $\mathcal{D}_{test}$ contains 20\% of the total number of samples of the smallest class in $\mathcal{D}$.
To train the GAN and the classifier we use and $D_{train}^{u}$, a uniformly balanced version of the original $\mathcal{D}_{train}$ obtained through the resampling block that will be explained in Section \ref{subsec:resample}.

After training the GAN, we use $G$ to create a synthetic uniformly balanced dataset $\mathcal{D}_{gen}^u$. Since $G$  performs conditional generation, to generate a uniformly balanced dataset we sample the conditional vectors $\bar{z}$ from $\mathcal{D}_{train}^{u}$. It is essential to mention that the generated dataset will follow the distribution of the dataset from which we sample the conditional vectors. For example, sampling them from $\mathcal{D}_{train}$ would imply generating a heavily unbalanced dataset. To obtain the TSTR score, we train a classifier on $\mathcal{D}_{gen}^u$ and evaluate its accuracy on a real dataset.
 
We compare the TSTR score to multiple TRTR scores, computed in a similar manner but using $\mathcal{D}_{train}^{u}$(or slightly modified versions of it) instead of $\mathcal{D}_{gen}^u$. This comparison is reasonable because the datasets used for evaluation ($\mathcal{D}_{val}$ and $\mathcal{D}_{test}$) are fixed and balanced by construction: their sampling process from $\mathcal{D}$ is designed to have the same amount of samples per class.  

\subsection{Data structure details}
\label{subsec:dsdetails}

Let $\phi_t$, $a$,\text{ and} $c$ denote the observation phases, amplitudes and classes of the light curves  respectively, our GAN's generator requires a sample $\bar{z} = [\phi_t, a, c]$ from the real dataset $\mathcal{D}_{train}$ and a sample $z\in\mathbb{R}^{\ell}\sim\mathcal{N}(0,I)$. The latent space dimensionality $\ell$ is set to 16 and 8 for the Catalina and ZTF datasets, respectively, obeying roughly the proportion between the light curve lengths of the datasets. Following a CGAN-like approach \citep{cgans2014}, the concatenation of $z$ and $\bar{z}$ is passed as an input to $G$ to generate synthetic samples. 

The conditional parameters are also inputs of D similarly concatenated with real or generated magnitudes. We create a tensor version of the conditional parameters for this concatenation to be viable. Let $\mathbf{a}$ and $\mathbf{c}$ be tensor versions of $a$ and $c$, and $L$ and $N$ denote the light curve length and number or classes of a dataset; we define $\mathbf{a} \in \mathbb{R}^L$ as a vector with value $a$ in all its components, and $\mathbf{c} \in \mathbb{R}^{L \times N}$ as a one-hot encoding of $c$, composed by $\mathbf{0}$'s and $\mathbf{1}$'s vectors, where $\{\mathbf{0}, \mathbf{1}\} \in \mathbb{R}^L$. The tensor version of $\bar{z}$ is $\mathbf{\bar{z}} = [\phi_t, \mathbf{a}, \mathbf{c}] \in \mathbb{R}^{L \times 2 + N}$.
The concatenation of $\mathbf{\bar{z}}$ and real or generated magnitudes will be the input of $D$, and will be dimensions $L \times  3+N$.

\subsection{Classifier details}
\label{subsec:clf}

To reduce the variance of the experiments, the classifier consists of an ensemble of 5 identical base-classifiers trained independently. The base-classifier is a CNN that receives the concatenation of the magnitudes $x$ and phases $\phi_t$ following the classification scheme in \citet{tcgan2018}. The input is forwarded through a set of convolution blocks that halve the temporal dimension, followed by dense layers. The network is trained using Adam optimizer \citep{adam2014} with $\alpha=0.0001, \beta_1=0.9, \beta_2=0.999$. Table \ref{table:clfarch} shows the detailed architecture of the base classifier. To compute all the feature-based metrics explained in Section \ref{subsubsec:featmetrics}, we use the output of the last convolution block of this base classifier, trained on each of the datasets separately.

\begin{table}[!htbp]
\centering
\normalsize
\caption{Classifier architecture. $L, \text{ and } N$ correspond to the light curve length and number of classes respectively and they vary depending on the selected dataset as mentioned in Section \ref{sec:data}. The fixed block parameters $p_s$ and $k_s$ stand for pool size and kernel size. Since the convolution blocks always halve the temporal dimension, we only specify their channel dimensions $c_{in}$ and $c_{out}$.}
\label{table:clfarch}
% \resizebox{\linewidth}{!}{%
    \begin{tabular}{l c}
    \hline \hline
    \multirow{2}{*}{Input} & $x \in \mathbb{R}^{L}$ \\
    {} &  $\phi \in \mathbb{R}^{L}$ \\
    Conv. Block & $2 \rightarrow 32$ \\
    Conv. Block & $32 \rightarrow 64$ \\
    Conv. Block & $64 \rightarrow 128$  \\
    Conv. Block & $128 \rightarrow 64$ \\
    Conv. Block & $64 \rightarrow  64$ \\
    Dense & $\ceil*{L/32} \times 64 \rightarrow 100$  \\
    BN, ReLU, Dropout & $100 \rightarrow 100$ \\
    Dense, Softmax & $100 \rightarrow N $ \\
    \hline \hline 
    \multicolumn{2}{c}{Convolution Block $(p_s=2, k_s=3, c_{in}, c_{out})$} \\
    \hline
    Block Input & $l_i \times c_{in}$ \\
    1-D Convolution, BN & $l_i \times c_{in} \rightarrow l_i \times c_{out}$ \\
    Max-pooling, ReLU & $l_i \times c_{out} \rightarrow \ceil*{l_i/2} \times c_{out}$  \\
    \end{tabular}
% }
\end{table}

\subsection{GAN details}

In addition to the original WGAN-GP formulation, we include additional regularization terms to Equation \ref{eq:ld} and \ref{eq:lg}. Following an AC-GAN-like approach \citep{acgans2017}, the output of $D$ has two components: $D_{rg} \in \mathbb{R}$ that tries to separate real from fake samples and $D_y \in \mathbb{R}^N$ that tries to predict the class of the input. Therefore, we add a cross-entropy regularization of real and generated samples to the discriminator loss. Also, to prevent the GAN equilibrium from happening in any arbitrary offset, we add a regularization term to prevent $D_{rg}(x_r)$ from drifting too far away from zero, as proposed in \citet{progan2018}. To the generator loss, we only add the cross-entropy regularization of generated samples. Consequently, the losses minimized in the proposed framework are:

\begin{eqnarray}
    \widetilde{L_D} &=& L_D + \xi (H_r + H_g) +  \epsilon \mathbb{E}[D_{rg}(x_r)^2] \\
    \widetilde{L_G} &=& L_G + \xi H_g
\end{eqnarray}

\noindent
where $H_r=H(y_r, D_y(x_r))$ and $H_g=H(y_g, D_y(x_g))$ correspond to the cross-entropy between the real labels and the discriminator predictions, $y_g$ are the real labels used to generate $x_g$, and $\xi=0.001$ and $\epsilon=1$ control the strength of each regularization term.

We perform $n_{disc}=5$ discriminator iterations per generator iteration, and train for 400K generator iterations using Adam optimizer with $\alpha=0.0001, \beta_1=0.5, \beta_2=0.9$. At training time, we compute the Exponential Moving Average (EMA, \citealt{ema2019}) with decay 0.999 for the generator weights, to be used when generating samples for evaluation. A full description of the GAN architecture is shown in Table \ref{tab:ganarch}.

On the one hand, $G$ receives the concatenation of the noise source $z$ and the conditional variables $\bar{z}$ as an input, and it forwards it through a dense layer followed by a set of strided deconvolutions that duplicate the temporal dimension of every block and simultaneously halving the number of channels (except for the last block). On the other hand, $D$ receives the concatenation of the magnitudes $x$ and the conditional tensor $\mathbf{\bar{z}}$, and it forwards it through a set of strided convolutions that halve the temporal dimension of every block and duplicate the number of channels, followed by a dense layer.

\begin{table*}[!htpb]
    \caption{GAN architecture. $\ell, L, \text{ and } N$ correspond to the latent space dimensionality, light curve length and number of classes respectively, which depend on the selected dataset as mentioned in Sections \ref{sec:data} and \ref{sec:meth}. The fixed block parameters $s$ and $k_s$ stand for stride and kernel size respectively, and $l_i$ represents the input length of the blocks. Since the convolution/deconvolution blocks always adjust the temporal dimension by a factor of 2, we only specify their channel dimensions $c_{in}$ and $c_{out}$.}
    \begin{minipage}{.49\textwidth}
        \centering
        \normalsize
        \begin{tabular}{lc}
            \multicolumn{2}{c}{(a) Generator} \\
            \hline \hline
            \multirow{2}{*}{Input} & $z \in \mathbb{R}^{\ell}$ \\
            {} &  $\bar{z} \in \mathbb{R}^{L+1+N}$ \\
            Dense, ReLU &  $\ell+(L+1+N) \rightarrow 4 \times 1024$\\
            Deconv. Block & $1024 \rightarrow 512$\\
            Deconv. Block & $512 \rightarrow 256$\\
            Deconv. Block & $256 \rightarrow 128$\\
            Deconv. Block & $128  \rightarrow 64$\\
            Deconv. Block & $64 \rightarrow 1$\\
            Tanh $\cdot s $  & $L \times 1$ \\ 
            \hline \hline 
            \multicolumn{2}{c}{Deconvolution Block $(s=2, k_s=5, c_{in}, c_{out})$} \\
            \hline
            Block Input & $l_i \times c_{in}$ \\
            1-D Deconvolution & $l_i \times c_{in} \rightarrow 2 l_i \times c_{out}$ \\
            ReLU & $2 l_i \times c_{out}$ \\
        \end{tabular}
    \end{minipage}%
    \begin{minipage}{.49\textwidth}
        \centering
        \normalsize
        \begin{tabular}{lc}
            \multicolumn{2}{c}{(b) Discriminator} \\
            \hline
            \hline
            \multirow{2}{*}{Input} & $x \in \mathbb{R}^{L \times 1}$ \\
            {} &  $\mathbf{\bar{z}} \in \mathbb{R}^{L \times (2+N)}$ \\
            Conv. Block & $1+(2+N) \rightarrow  64$\\
            Conv. Block & $64 \rightarrow 128$\\
            Conv. Block & $128 \rightarrow 256$\\
            Conv. Block & $256 \rightarrow 512$\\
            Conv. Block & $512 \rightarrow 1024 $\\
            Dense & $\ceil*{L/32} \times 1024 \rightarrow N+1$  \\
            \hline \hline
            \multicolumn{2}{c}{Convolution Block $(s=2, k_s=5, c_{in}, c_{out})$} \\
            \hline 
            Block Input & $l_i \times c_{in}$\\
            1-D Convolution & $l_i \times c_{in} \rightarrow \ceil*{l_i/2} \times c_{out} $ \\
            LeakyReLU & $\ceil*{l_i/2} \times c_{out} $ \\
        \end{tabular}
    \end{minipage}%
    \label{tab:ganarch}
\end{table*}

\subsection{Preliminary experiment: The $u$-GAN}
\label{subsec:uGAN}

\begin{figure*}[!htbp]
\gridline{\fig{
               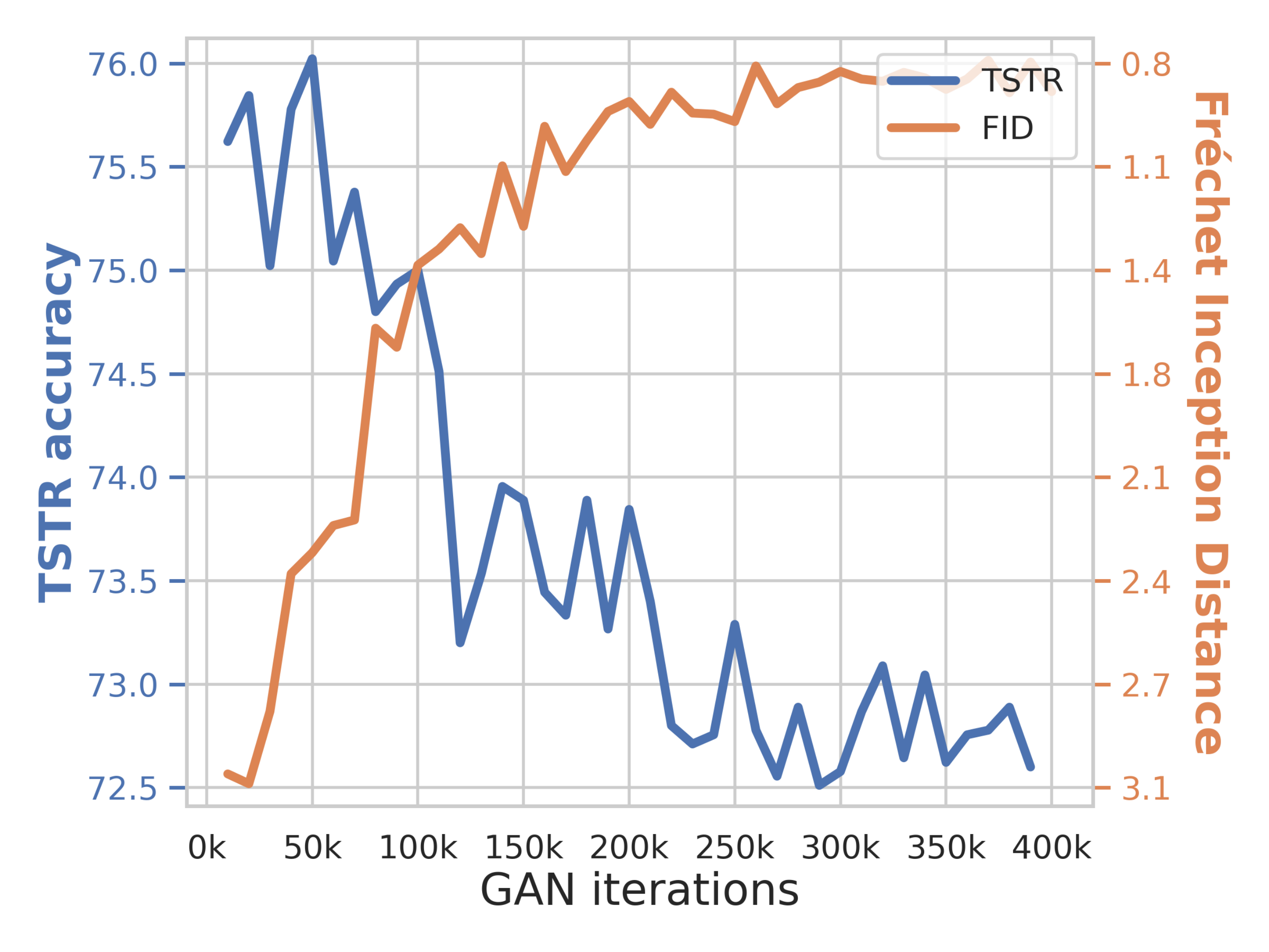}{0.49\textwidth}{(a) Catalina}
          \fig{
               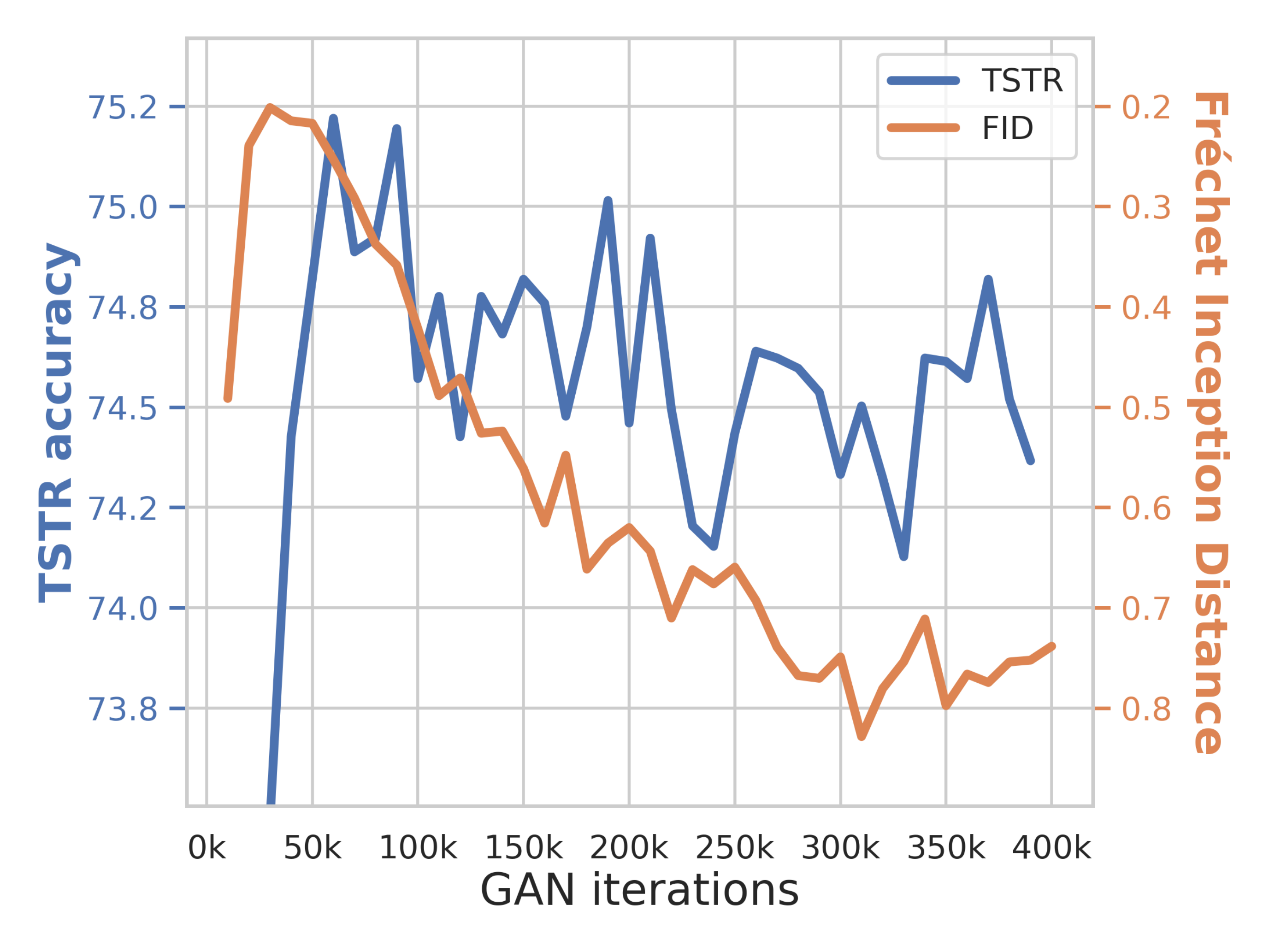}{0.49\textwidth}{(b) ZTF}
         }
\caption{Evolution of the validation TSTR accuracy and FID over the course of GAN training for the different datasets. Both scores were computed every $10k$ iterations of a single GAN model. The computation of the FID was done with $50k$ generated samples divided into 10 batches and the entire real dataset, as suggested in \citet{ttur2017}.}
\label{fig:tstr_fid_vs_it}
\end{figure*}

With all details and parameters provided in the above sections, we perform a preliminary experiment using $\mathcal{D}_{train}^u$ -- the uniformly balanced version of $\mathcal{D}_{train}$ -- as the GAN training set, to then generate $\mathcal{D}_{gen}^u$ and obtain the TSTR accuracy scores.
This GAN setup will be referred to as the ``$u$-GAN". 

It is worth mentioning that this setup is the standard approach when training machine learning algorithms, where $\mathcal{D}_{train}^u$ is usually preferred over $\mathcal{D}_{train}$ because it reduces the biases towards the most populated classes, induced by the highly imbalanced class distribution of $\mathcal{D}_{train}$. 

% Talk about the variability of the TSTR and why we show a curve instead of a point. 
The first finding of performing this preliminary experiment is that the TSTR accuracy score can vary significantly depending on how long we train the GAN. For this reason, we analyze the behavior of different GAN models throughout the training process to find an adequate criterion for model selection. Figure \ref{fig:tstr_fid_vs_it} shows the evolution of the validation TSTR accuracies and FID scores every $10k$ iterations. Since computing TSTR accuracies involves training multiple classifiers, evaluating this score more frequently is unfeasible. 

The preliminary experiment shown in Figure \ref{fig:tstr_fid_vs_it} raises two major concerns that will be addressed in the following sections: 

\begin{enumerate}[label=\alph*)]
    \item \label{item:1} The TSTR accuracy reaches an optimal value early in the GAN training and then decreases consistently, coinciding with the GAN overfitting phenomenon explained in Section \ref{subsec:of}. 
    
    \item \label{item:2} The FID -- the standard metric for evaluating GANs -- cannot always measure the drop in sample quality reflected in the TSTR accuracy curve, as shown in Figure \ref{fig:tstr_fid_vs_it}a.
\end{enumerate}

% What is a) and why is it a problem
The behavior detailed in \ref{item:1} can be understood as follows: in a balanced dataset such as $\mathcal{D}_{train}^{u}$, overfitting is not only strongly influenced by the limited amount of training samples, but it also is exacerbated by the amount of imbalance of the original class distribution of $\mathcal{D}_{train}$. As the imbalance grows, samples in the minority classes need to be excessively repeated in order to equate the number of samples in the majority classes, resulting in quick GAN overfitting caused by $D$ learning fast how samples of the minority classes look. The rapid decay in validation TSTR accuracy is problematic considering that we need to compute this metric every $10k$ iterations. Hence, the best model selected by this metric could be sub-optimal if the decay occurs suddenly, which motivates the proposed \textbf{resampling block} explained in Section \ref{subsec:resample}.
            
% What is b) and why does it make sense            
The discrepancy described in \ref{item:2}, although undesirable, is not surprising; it was also reported in \citet{cas2019}, and it is completely plausible considering the limitations of FID related to mode dropping and mode inventing mentioned in Section \ref{subsubsec:featmetrics}. These two phenomena can drastically affect how $P_g$ relates to $P_r$ and thus affect the TSTR accuracy without being reflected in the FID, which suggests that FID is not always reliable in the presence of highly unbalanced datasets, and motivates the proposed $\mathbf{\mathcal{G}}$ \textbf{-score} for model selection explained in Section \ref{subsec:modelsel}.

% ALSO: 
% Two aspects that could cause the discrepancy in \ref{item:2} could be the normality assumptions of the FID score which could be inappropriate with small imbalanced datasets, and the lack of expressiveness of considering only the first two statistical moments when computing the FID score as suggested by \citet{ttur2017}.

\subsection{Resampling block}
\label{subsec:resample}
% Motivation

Motivated by the rapid GAN overfitting shown in Figure \ref{fig:tstr_fid_vs_it}, we propose a resampling operation that can successfully delay the occurrence of this behavior. 

% General description
The resampling operation consists of continuously drawing samples from the $N$ classes of a dataset $\mathcal{D}$, to modify its class distribution.  
% Dig into more details
Let $S$ be the number of samples of $\mathcal{D}$. We start by splitting $\mathcal{D}$ into $N$ sub-datasets $\{\mathcal{D}_{i}\}_{i=1}^{N}$ of size $\{S_{i}\}_{i=1}^{N}$, where each dataset $\mathcal{D}_{i}$ only contains samples from the $i$-th class. From each sub-dataset, we draw without replacement until there are no samples left, then $\mathcal{D}_i$ is shuffled and the sampling process continues. 

The goal of this operation is to modify the class distribution of $\mathcal{D}$ by controlling the probability $p_i$ of drawing a sample from each $\mathcal{D}_{i}$. The resampling block serves as a generalization of the uniform balancing operation by extending the target class distribution to non-uniform distributions. To illustrate this clearly, we describe two edge cases. On the one hand, we could leave the original class distribution unbalanced by setting $p_i =S_i / S$, in which case the resampling block does not affect the class distribution, and it would be equivalent to a ``shuffle and repeat" operation. On the other hand, we could obtain the balanced version of $\mathcal{D}$ by simply setting $p_i= 1/N$, which is how we get  $\mathcal{D}_{train}^{u}$ from $\mathcal{D}_{train}$.

\begin{figure*}[!htbp]
\gridline{\fig{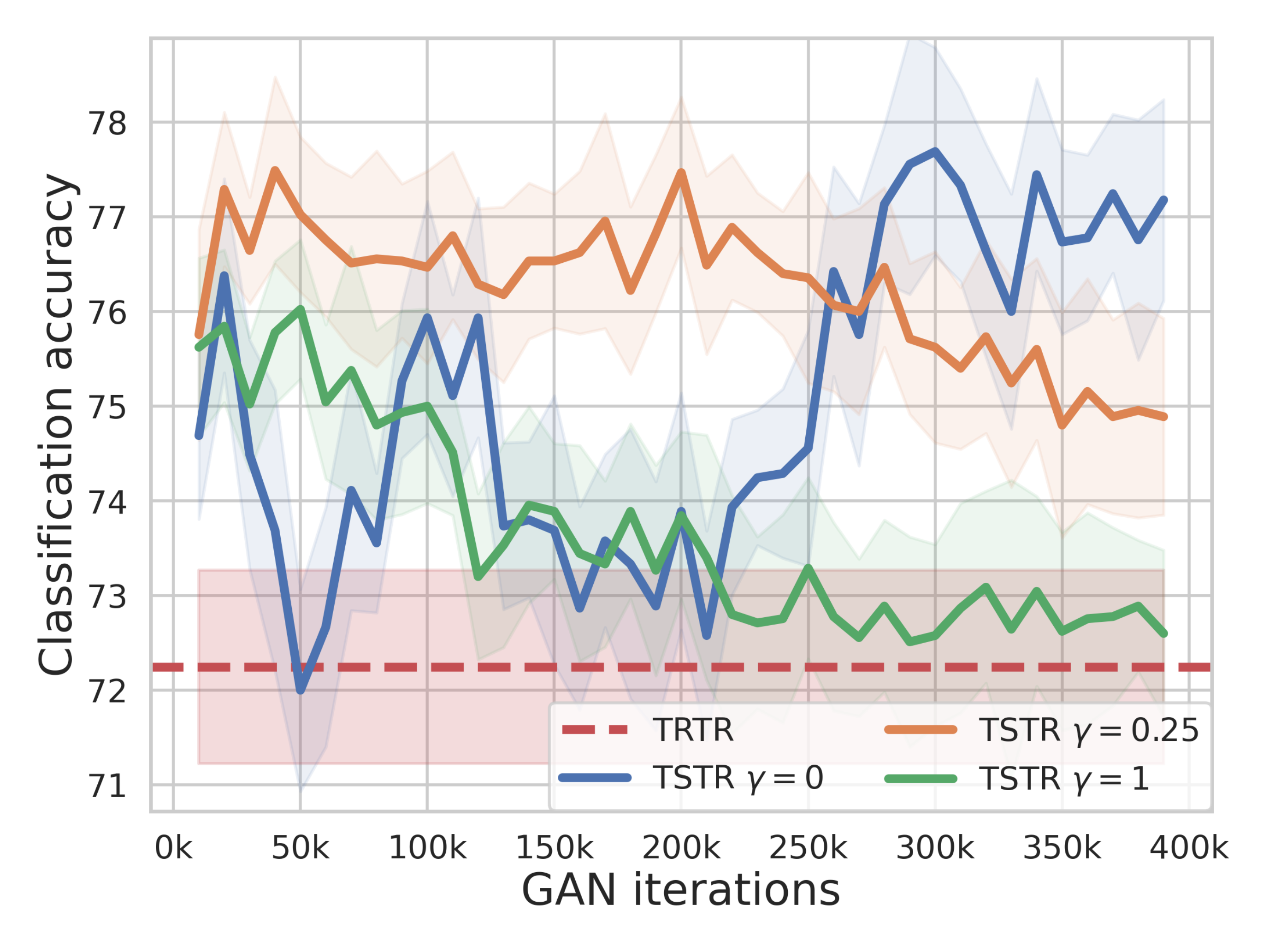}{0.49\textwidth}{(a) Catalina}
          \fig{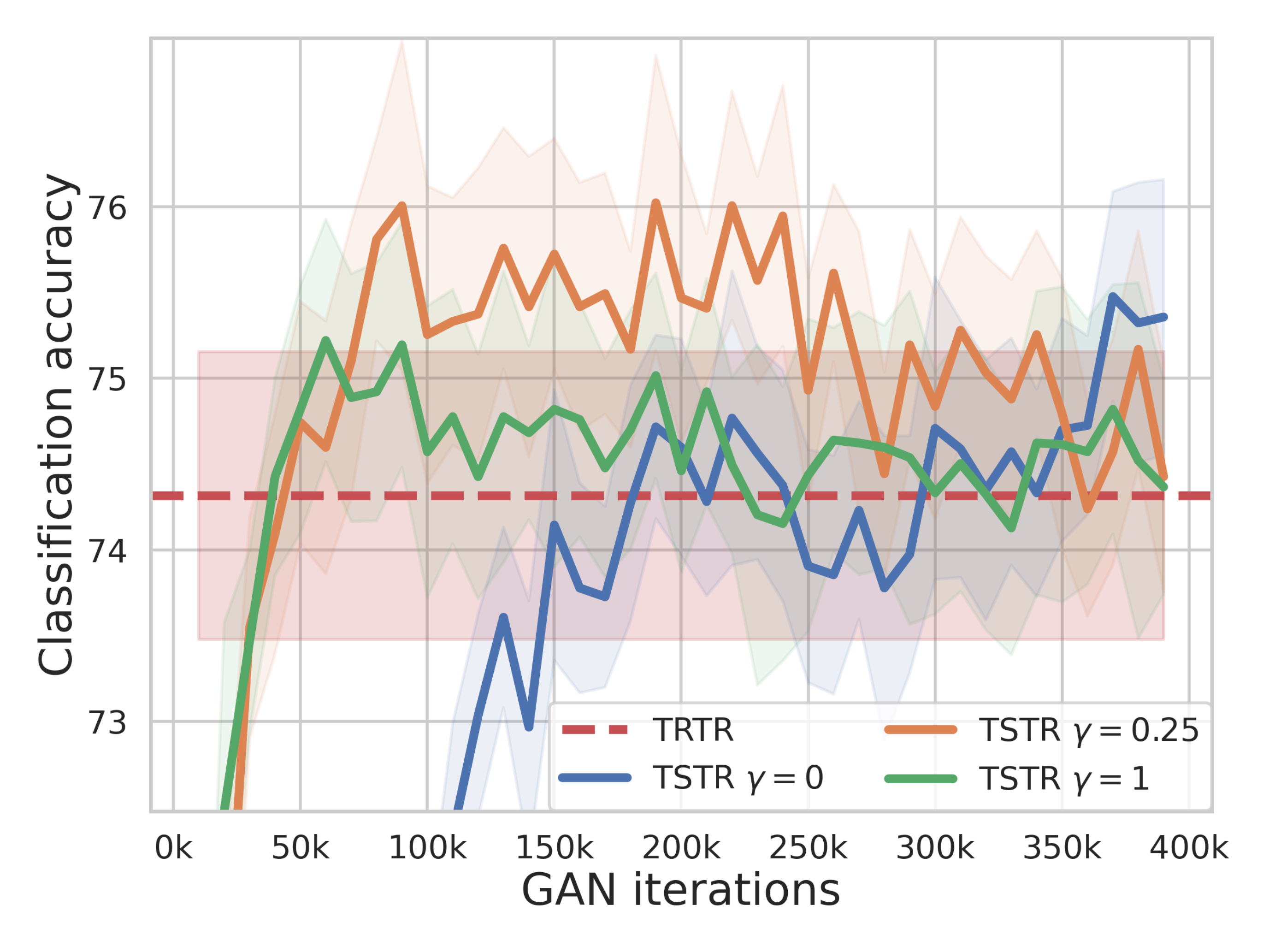}{0.49\textwidth}{(b) ZTF}}
\caption{Evolution of TSTR accuracy over the course of GAN training for different values of $\gamma$. The figure shows mean $\pm$ standard deviation over 15 independent runs of the classifier and a single GAN model. The computation of both metrics was done every $10k$ GAN iterations}
\label{fig:delta_tstr_vs_it}
\end{figure*}

Apart from these two scenarios, we could also generate any dataset $\mathcal{D}^\gamma$ whose class distribution lies "in between" that of $\mathcal{D}$ and $\mathcal{D}^{u}$, created by linearly interpolating between the aforementioned probabilities:

\begin{equation}
    p_i = \gamma \left(\frac{1}{N}\right) + \left( 1-\gamma \right) \frac{S_i}{S}, \text{ where } 0< \gamma < 1 
\end{equation}

\noindent where the two edge cases can be recovered with $\gamma=0$ for the imbalanced $\mathcal{D}$, and $\gamma=1$ for the balanced $\mathcal{D}^{u}$.
By using the proposed $\gamma$-resampling we are able to control the overfitting speed of the model, as shown in Figure \ref{fig:delta_tstr_vs_it}. Training a GAN with $\mathcal{D}^{u}(\gamma=1)$ implies that all the samples from the minority classes are rapidly shown to the model, leading to fast overfitting. On the other hand, using $\mathcal{D}(\gamma=0)$ implies that training batches rarely contain a sample from the minority classes (1 every 230 samples will be cepheids of the Catalina dataset, roughly 1 cepheid every 4 batches), avoiding fast overfitting but inducing slow and unstable training. Training with $\mathcal{D}^{\gamma}(0<\gamma<1)$ allows a reasonable learning pace without overfitting rapidly, as shown in Figure \ref{fig:delta_tstr_vs_it} for $\gamma=0.25$. A model trained with $\mathcal{D}^{\gamma}$ will be referred to as the ``$\gamma$-GAN"

\subsection{Model selection: The $\mathcal{G}$-score}
\label{subsec:modelsel}

% Motivation: accuracies are expensive 
As mentioned in Section \ref{subsec:uGAN}, the behavior of TSTR accuracies shown in Figure \ref{fig:tstr_fid_vs_it} evidences the need for a criterion to choose an adequate $G$. While using the validation TSTR accuracy for model selection might look appropriate, doing so involves training new classifiers for every candidate of $G$, an operation that becomes computationally expensive. The problem then lies in finding a fast-to-compute metric that correlates with the TSTR accuracy (and implicitly with the quality of the generated samples). 

The natural option for this metric would be FID, but as also shown in Figure \ref{fig:tstr_fid_vs_it}a, it fails to measure the decrease in quality of the generated samples reflected in the TSTR accuracy curve. Additionally, since FID is only a measure of the distance between $P_g$ and $P_r$, it cannot differentiate between the fidelity and diversity of the generated samples \citep{dc2020}, and it provides an arbitrarily weighted average between them.

\begin{figure}[!htbp]
\includegraphics[width=.48\textwidth]{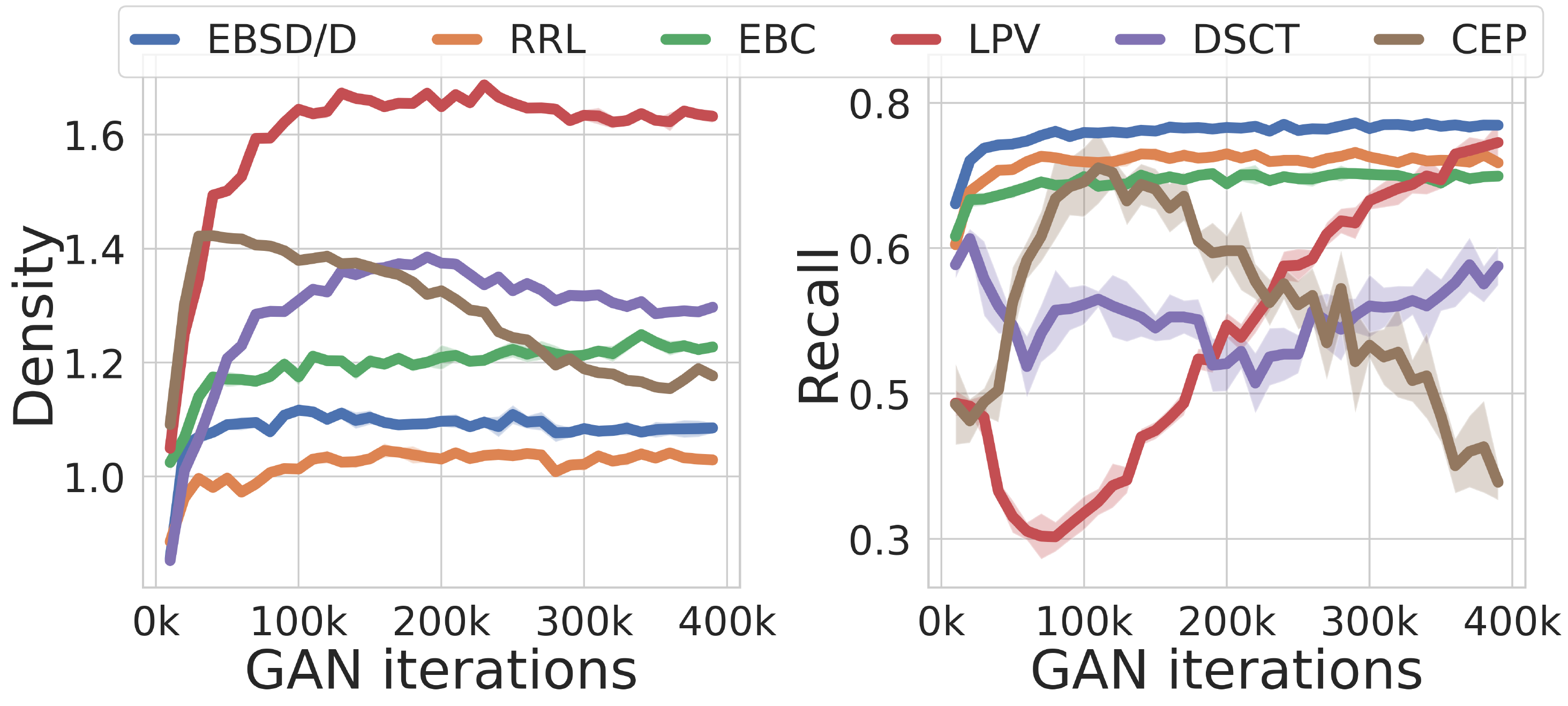}
\caption{Class density and recall metrics of the Catalina dataset.}
\label{fig:DRCatalina}
\end{figure}

% Explain the G-score
As an alternative, we propose a metric that leverages equally two measures of fidelity and diversity: density\textbf{(D)} and recall\textbf{(R)}. Figure \ref{fig:DRCatalina} shows the results of computing the per-class density and recall metric for the Catalina dataset.

\begin{figure}[!htbp]
\includegraphics[width=.48\textwidth]{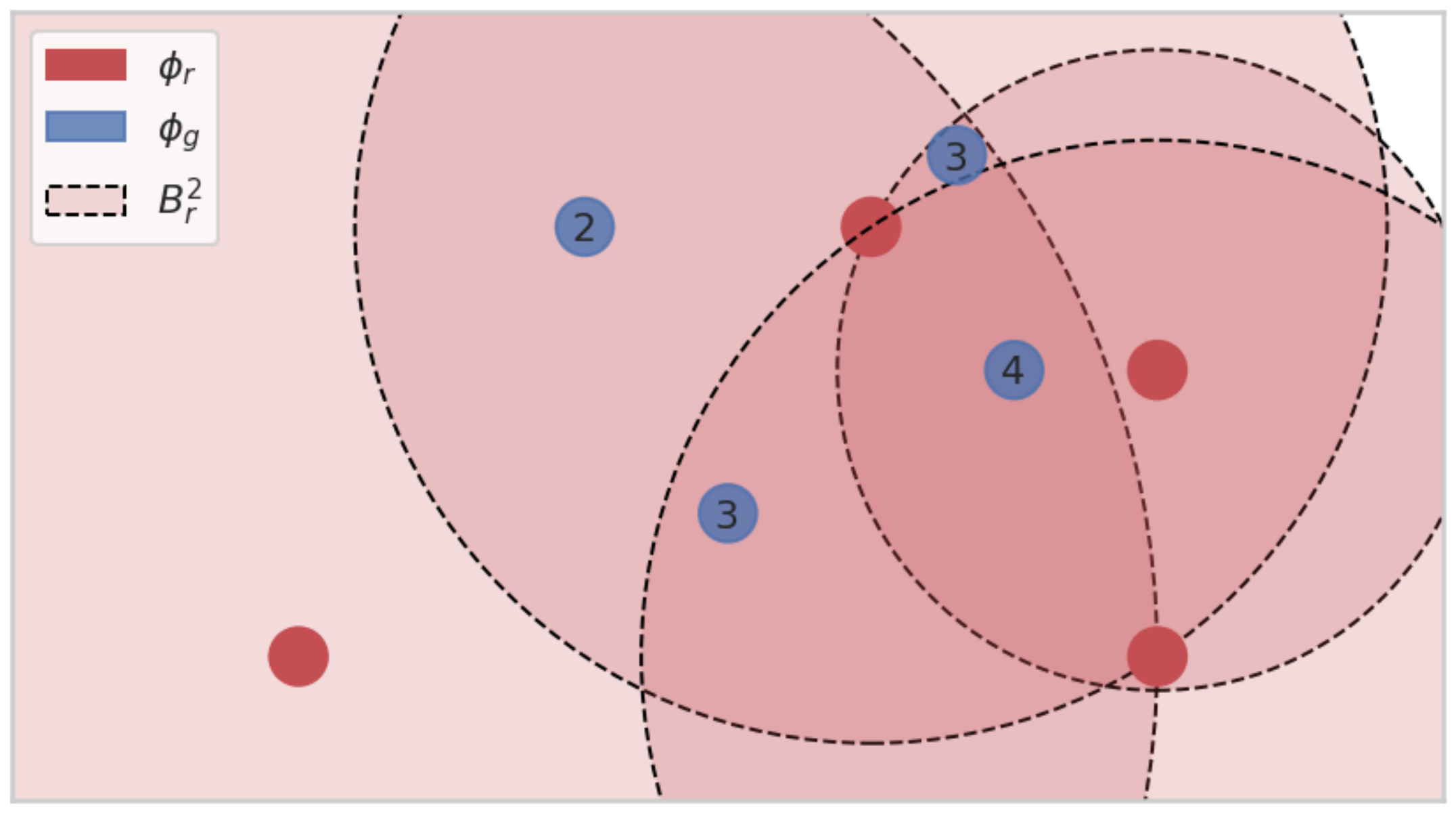}
\caption{Two-dimensional scenario that illustrates a case in which \textbf{D} is not bounded by 1. The dashed lines show the regions $B_r^2$: circles around the real feature samples $\phi_r$, with radii equal to the distance to their second nearest neighbors. The numbers inside each sample $\phi_g$ denote the number of circles that enclose the sample.}
\label{fig:Manifolds}
\end{figure}

% Mention the bias of the Density metric
The fact that \textbf{D} values are not bounded by one is consistent with the formula presented in Equation \ref{eq:density} and can happen if points in the generated manifold in average belong to more than $K$ balls of the real manifold, which is probably caused by the over-estimation of the real manifold mentioned in Section \ref{subsubsec:featmetrics}, due to sparse feature spaces. An illustration of this situation is shown in Figure \ref{fig:Manifolds}, where the sparsity in the real distribution causes that the generated samples in average belong to more than $K=2$ balls, leading to $\mathbf{D} = \frac{1}{4}(\frac{2}{2} + \frac{3}{2} + \frac{4}{2} + \frac{3}{2}) = 1.5 $. Additionally, if we reduce the sparsity of the real distribution by removing the furthest sample (bottom left), we get $\mathbf{D} = \frac{1}{4}(\frac{1}{2} + \frac{2}{2} + \frac{3}{2} + \frac{2}{2}) = 1$.

% Explained the selected approach to obtain the G-score
Since \textbf{R} is bounded between 0 and 1 by definition, the unbounded behavior of \textbf{D} is undesirable because it privileges \textbf{D} over \textbf{R} in any mean we compute between them. In addition, we find that \textbf{D} also presents a clear bias towards the less populated classes. To overcome these problems, we perform a per-class min-max normalization to \textbf{D} and \textbf{R} according to Equation 3.

\begin{eqnarray}
    \label{eq: min_max}
    \mathbf{D}_i' = \frac{\mathbf{D}_i - \mathbf{D}_i^{min}}{\mathbf{D}_i^{max} - \mathbf{D}_i^{min}} \\
    \mathbf{R}_i' = \frac{\mathbf{R}_i - \mathbf{R}_i^{min}}{\mathbf{R}_i^{max} - \mathbf{R}_i^{min}}
\end{eqnarray}

\noindent
where the subscript $(\cdot)_i$ denotes score of the $i$-th class, and the superscripts $(\cdot)^{min, max}$ denote the minimum and maximum score of the class respectively.

After the class scores are normalized, we combine them in an equally weighted $F$-score described in Equation \ref{eq:F}. Finally, considering that we are equally interested in the different classes, the $\mathcal{G}$-score is obtained by computing the balanced $F$-score (macro $F$-score), as shown in Equation \ref{eq:G}.  
 
\begin{eqnarray}
F_{i}  &=& \frac{2\mathbf{D}_i'  \mathbf{R}_i' }{\mathbf{D}_i' + \mathbf{R}_i' } \label{eq:F} \\
\mathcal{G}\text{-score} &=&  \frac{1}{N}\sum_{i}F_{i} \label{eq:G}
\end{eqnarray}

When computing the balanced $F$-score, we prefer the mean of the class $F$-scores over the $F$-score of the class means intending to weight equally majority and minority classes, as suggested by \citet{macrof2019}.

\begin{figure*}[!htbp]
\gridline{\fig{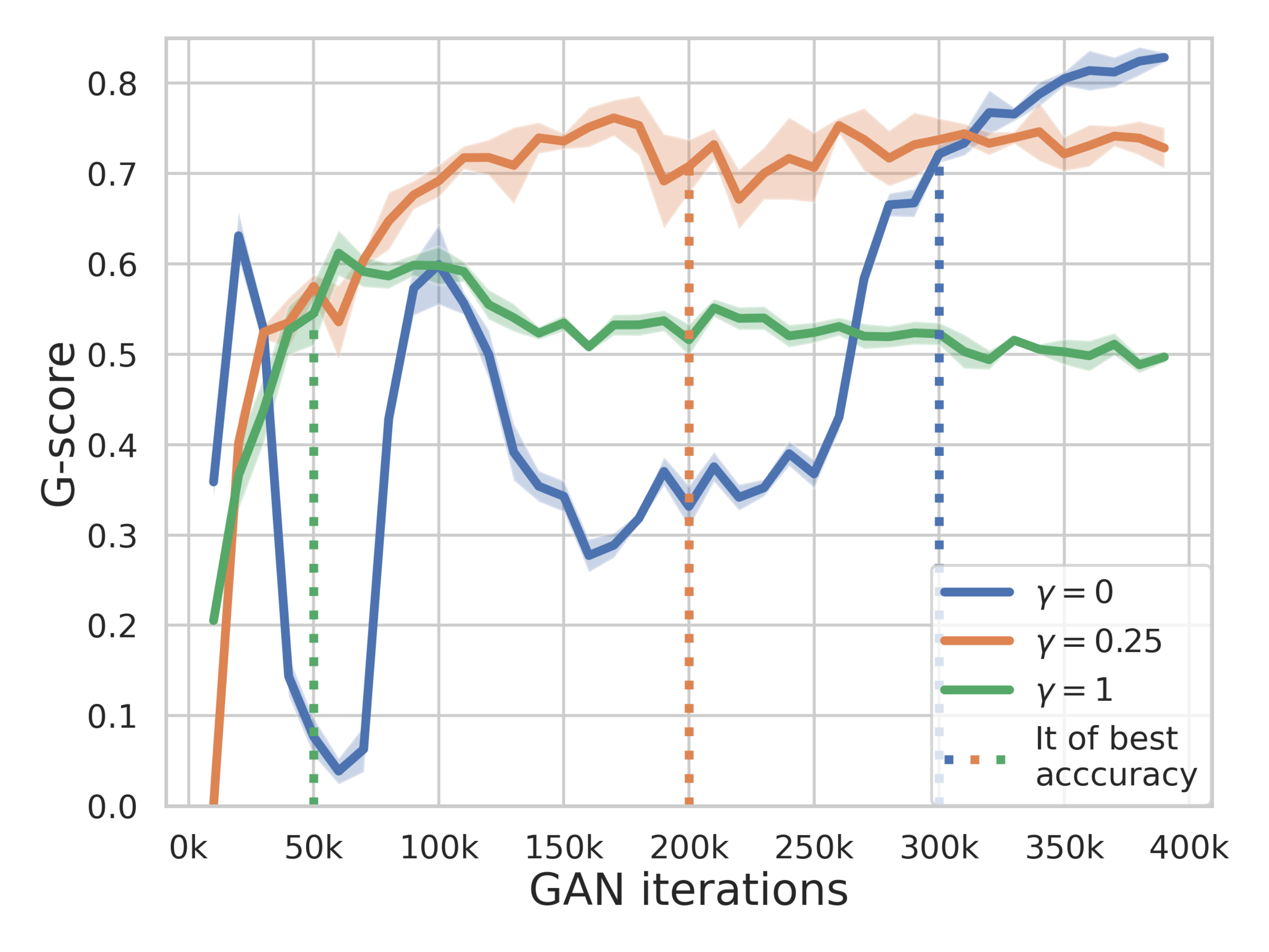}{0.49\textwidth}{(a) Catalina}
          \fig{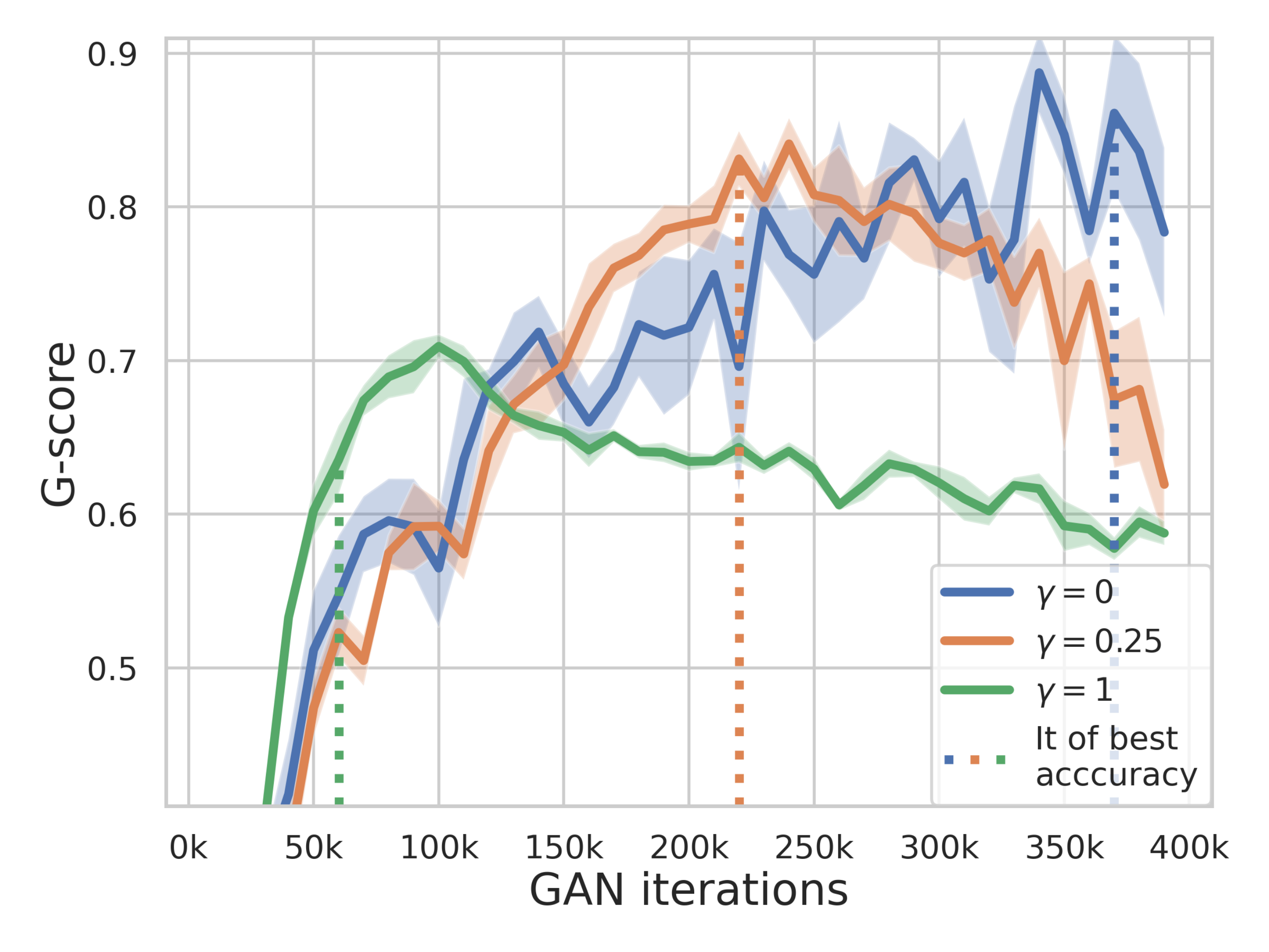}{0.49\textwidth}{(b) ZTF}}
\caption{Evolution of the $\mathcal{G}$-score for different $\gamma$ values over the course of GAN training for the different datasets. Each curve shows mean $\pm$ standard deviation over 5 computations of the metrics, for a single GAN model }

\label{fig:Gscore_vs_it}
\end{figure*}

The results of computing the $\mathcal{G}$-score for multiple GANs trained with different values of $\gamma$ are shown in Figure \ref{fig:Gscore_vs_it}. As it can be seen, the $\mathcal{G}$-score curves and validation accuracy curves from Figure \ref{fig:delta_tstr_vs_it} seem to have a high correlation, which becomes more evident when analyzing the $\gamma=0$ curve for the Catalina dataset.

\subsection{Baselines}
\label{subsec:baselines}

To evaluate our generated datasets in the classification task, we compare the TSTR classification accuracies to multiple baselines. These baselines consist of TRTR classification accuracy scores when training in augmented real datasets. It is worth mentioning that the training sets used to compute the scores are all \textbf{balanced datasets}, either GAN-generated (TSTR) or real-augmented (TRTR).

Acknowledging the heteroscedastic behavior of astronomical data, we do not consider jittering as a suitable operation for the problem. Additionally, we discard utilizing window-slicing techniques since our convolutional architectures work on pre-processed time-series with a fixed number of observations. Consequently, our augmentation methods consist of oversampling and different window-warping-based operations.
\subsubsection{Oversampling}

The oversampling augmentation corresponds to generating the balanced dataset $\mathcal{D}_{train}^{u}$ by repeating samples from the original dataset $\mathcal{D}_{train}$, using the resampling block described in Section \ref{subsec:resample}.

\subsubsection{Window-warping}

Let $x(t)$ be a continuous signal sampled at times $t$. The window-warping operation starts by selecting a random time window delimited by the values $\left[t_1,t_2\right]$, where all the times $t_w$ in the window satisfy $t_1 \leq t_w \leq t_2$. The warping operation expands or contracts the signal by scaling the variations $\Delta t$ in $t_w$ and shifting the times $t > t_2$ accordingly, altering the time-series' length. 

Since we work with folded light curves in phase space, window-warping expansion could be incongruous with the fact that the phase space has an upper bound of 1. Consequently, we derive a new transformation to avoid such incongruence: \textbf{soft window-warping}.
%\pp{I am missing the big idea here. Please expand and explain what is the modification you do and why is it a problem with the bounds} 
%\ggj{Added the shifting phrase, and changed the bound part to be more specific. The modifications are explained below}

\subsubsection{Soft window-warping}

% Soft window time-warping
We preserve the core idea of window-warping by designing expansions and contractions that do not increase the time-series' length. Given a random window, we formulate the problem as finding a mapping $t_w \mapsto f(t_w)$ such that the length of the transformed window is at most that of the original, this is $f(t_1)\geq t_1$, $f(t_2)\leq t_2$. We believe that expansions and contractions should be naturally performed with respect to the center of the window, expanding from the center to the limits and contracting from the limits to the center. 

A mapping that meets these requirements is:  
\begin{eqnarray}
    f(t_w) &=& a + b \cdot \tanh{(k(t_w-c))}   \\
    a &=& c = (t_1+t_2)/2 \nonumber \\
    b &=& (t_2-t_1)/2 \nonumber
\label{eq:soft-tw}
\end{eqnarray}

\noindent where the values of $a,b \text{ and } c$ are determined by the desired behavior with respect to the center of the window. The constant $k$ is randomly sampled in the interval $\left[\frac{1}{2a},\frac{2}{a}\right]$ and it modulates the strength of the expansions or contractions by modifying the saturation degree of the $\tanh{(\cdot)}$, producing expansions when saturated and contractions otherwise.

% How to extend it for magnitudes
Even though the proposed transformation is designed to be applied across the time axis, it can be easily extended to the signal axis by noting that since the time intervals are monotonous, $t_1, t_2$ are the minimum and maximum values in the window respectively. Hence, the natural extension to the signal axis is:
\begin{eqnarray}
\label{eq:soft-mg}
    f(x_w) &=& a + b \cdot \tanh{(k(x_w-c))} \\
    m_1 &=& \min\limits_{t \in t_w}{x(t)} \nonumber \\
    m_2 &=& \max\limits_{t \in t_w}{x(t)} \nonumber \\
    a &=& c = (m_1+m_2)/2 \nonumber \\
    b &=& (m_2-m_1)/2 \nonumber 
\end{eqnarray}

\begin{figure}[!htbp]
\includegraphics[width=.49\textwidth]{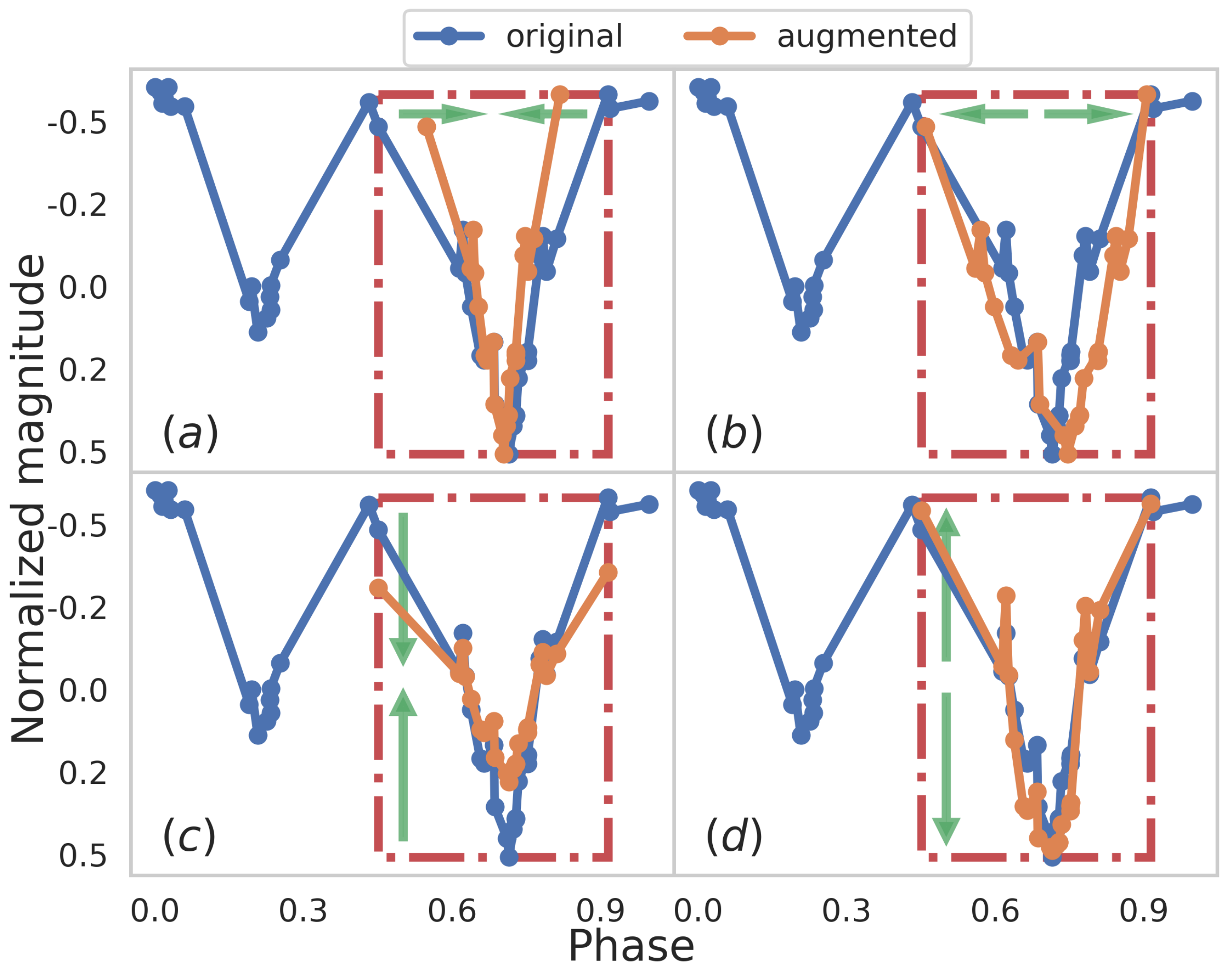}
\caption{Examples of the soft window-warping transformations for an eclipsing binary of the ZTF dataset. (a) Soft time-warping contraction. (b) Soft time-warping expansion. (c) Soft magnitude-warping contraction. (d) Soft magnitude-warping expansion.}
\label{fig:SoftWarping}
\end{figure}

When applying these transformations to our astronomical light curves, we consider the signal axis as the magnitude axis, and the time axis as the phase axis. These two transformations referred to as \textbf{soft time-warping} and \textbf{soft magnitude-warping}, are illustrated in Figure \ref{fig:SoftWarping}. The result of simultaneously applying these two transformations will be referred to as \textbf{soft mixed-warping}.

\section{Results}
\label{sec:results}

\subsection{Generated Samples}

\begin{figure*}[!htbp]
\centering
\includegraphics[width=\textwidth]{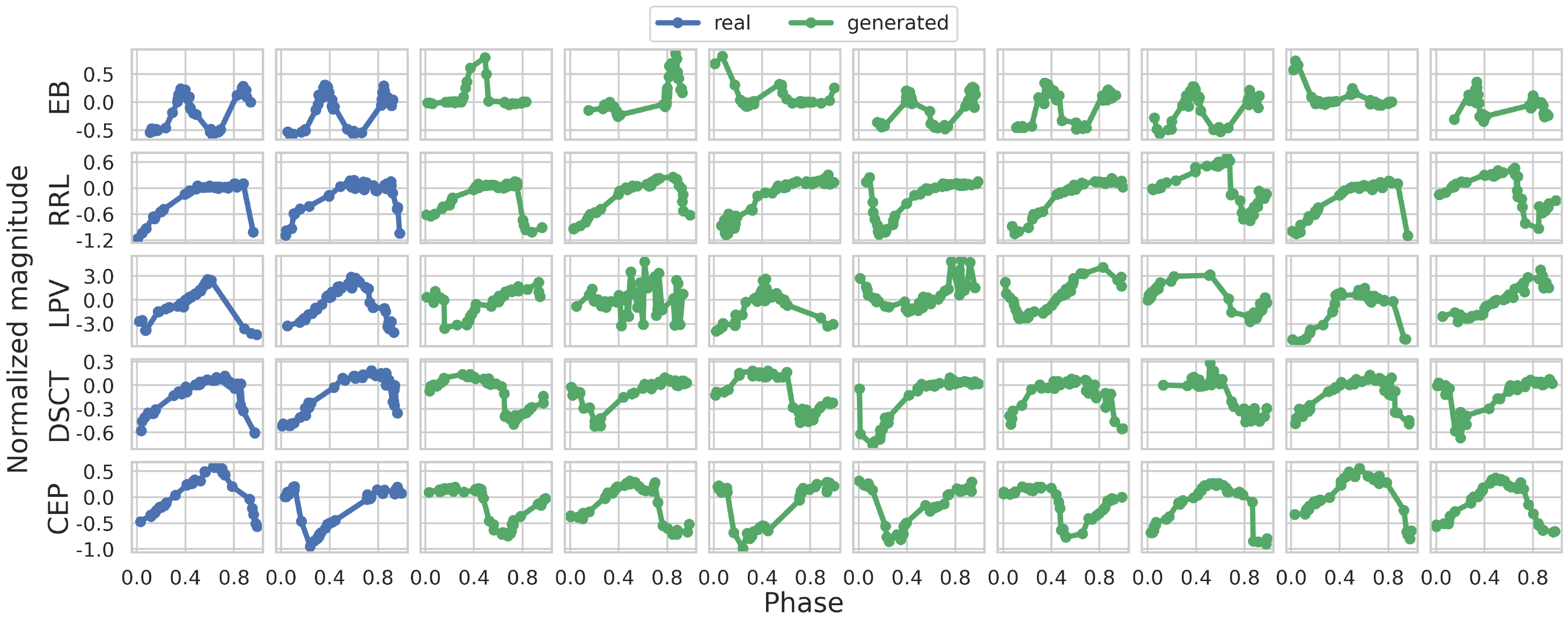}
\caption{Real and generated light curves of the ZTF dataset. To produce the synthetic curves in green, we perform conditional generation with the attributes (phases, class and amplitude) of the real curves in blue.}
\label{fig:GenSamples}
\end{figure*}

Figure \ref{fig:GenSamples} shows some samples of the GAN-generated light curves. The conditional vector $\bar{z}$ used to generate these samples considers phases, amplitudes, and classes of the real data shown in the first two columns. Accordingly, and as it can be seen, most of the generated samples preserve the real class and amplitude. It is worth mentioning that although some generated samples present normal fluctuations in phase and magnitude with respect to the real ones, there are also samples that do not look plausible (see Figure \ref{fig:RealismZTF}), which could be attributed to the lack of truncation techniques or any type of filtering to improve the fidelity of the generated samples, which we address in Section \ref{sec:discussion}.

\subsection{Classification}

The classification accuracies obtained by using different training sets are shown in Table \ref{table:AccResults}. The first four rows show TRTR classification results when training on real data that has been augmented with the random transformations described in Section \ref{subsec:baselines}. The soft-warping transformations (rows B-D) are applied to the dataset previously balanced by oversampling. The last four rows show TSTR classification results when training on GAN-generated data, comparing the proposed $\gamma$-resampling for GAN training ($\gamma$-GAN) against uniform resampling ($u$-GAN), and the proposed $\mathcal{G}$-score for model selection against the validation accuracy criterion. 

% Describe TRTR results 
As Table \ref{table:AccResults} shows, none of the soft-warping transformations achieves statistically significant differences with respect to the oversampling baseline (row A). 

% All GANs work
On the other hand, the benefits of using generative models are clear. Both GAN models achieve significant improvements with respect to the oversampling baseline, either using the validation accuracy criterion or the $\mathcal{G}$-score criterion for model selection.

% Uniform vs gamma 
We can also notice that using the $\gamma$-resampling can be beneficial in comparison to using the uniform approach. For both datasets, the minimum TSTR classification accuracy corresponds to the $u$-GAN (E for Catalina and F for ZTF), while the maximum corresponds to the $\gamma$-GAN (H for both datasets). Furthermore, for each dataset, the best TSTR accuracy is always significantly better than the worst. 

% Val accuray vs G-Score 
Regarding the model selection criteria, the $\mathcal{G}$-score shows to be an effective criterion, achieving accuracies that are at least statistically equivalent to the ones obtained by the computationally expensive validation accuracy criterion. Furthermore, it can sometimes obtain significantly better results, as shown in the ZTF dataset by the $\gamma$-GAN model.

% gamma + G-score nails it
Interestingly, the combination of the proposed $\gamma$-GAN + $\mathcal{G}$-score obtains the best classification accuracies overall, statistically outperforming all existing methods for ZTF dataset, and all but one ($\gamma$-GAN + val. accuracy) for the case of the Catalina dataset.

\begin{table*}[!htbp]
\centering
% \normalsize
\caption{Classification accuracy of the different augmentation methods on test datasets. For each method, we report the mean and standard deviation calculated over 15 independent runs. We also report the p-value of the two-sided Welch's tests between each method (rows) and the baselines shown with capital letters in the columns.}
\label{table:AccResults}
% \resizebox{\linewidth}{!}{%
% \makebox[\textwidth][c]{
    \begin{tabular}{cc c ccccc| ccccc} %2 1 5 5 
    \hline \hline
    {}&{} & \multirow{3}{*}{Method} & \multicolumn{5}{c}{Catalina} & \multicolumn{5}{c}{ZTF} \\ %\cline{2-5}
    {}&{} & {} & {Accuracy} & \multicolumn{4}{c|}{\textit{p} value} & {Accuracy} & \multicolumn{4}{c}{\textit{p} value} \\      
    {}&{} & {} & { [\%]}&\multicolumn{4}{c|}{A} & {[\%]}&\multicolumn{4}{c}{A} \\ \hline
    \multicolumn{1}{c|}{\multirow{4}{*}{\rotatebox{90}{TRTR}}}&{A} & {Oversampling} & {73.44$\pm$1.22}&\multicolumn{4}{c|}{} & {72.61$\pm$0.69}&\multicolumn{4}{c}{}\\
    \multicolumn{1}{c|}{}&{B} & {Soft time-warping}&{74.06$\pm$1.04}&\multicolumn{4}{c|}{.145} & {72.69$\pm$0.99}&\multicolumn{4}{c}{.786} \\
    \multicolumn{1}{c|}{}&{C} & {Soft mag-warping}&{73.64$\pm$1.79}&\multicolumn{4}{c|}{.723} & {72.45$\pm$0.70}&\multicolumn{4}{c}{.533} \\
    \multicolumn{1}{c|}{}&{D} & {Soft mixed-warping}&{73.82$\pm$1.50}&\multicolumn{4}{c|}{.452} & {72.53$\pm$0.69}&\multicolumn{4}{c}{.753} \\ \hline
    \multicolumn{1}{c|}{}&{} & {$u$-GAN}  & {}&{A}&{E}&{F}&{G} & {}&{A}&{E}&{F}&{G} \\
    \multicolumn{1}{c|}{\multirow{4}{*}{\rotatebox{90}{TSTR}}}&{E} & {Val Acc} & {75.97$\pm$0.94}&{$<$.001}&{}&{}&{}  &{74.17$\pm$0.62}&{$<$.001}&{}&{}&{}\\
    \multicolumn{1}{c|}{}&{F} &{$\mathcal{G}$-score} & {76.28$\pm$0.74}&{$<$.001}&{.324}&{}&{} &{73.79$\pm$0.50}&{$<$.001}&\multicolumn{1}{C{1.3}}{.075}&{}&{} \\
    \cline{2-13}
    \multicolumn{1}{c|}{} & {} & {$\mathbf{\gamma}$-\textbf{GAN}} &\multicolumn{5}{c|}{}&\multicolumn{5}{c}{} \\
    \multicolumn{1}{c|}{}&{G} & {Val Acc} & {76.86$\pm$1.09}&{$<$.001}&{.024}&{.102}&{} & {74.37$\pm$0.51}&{$<$.001}&\multicolumn{1}{C{1.3}}{.342}&\multicolumn{1}{C{1.3}}{0.003}&{}  \\ 
    \multicolumn{1}{c|}{}&{H} & {$\mathcal{G}$-score} & {$\mathbf{76.97\pm0.79}$}&{\textless.001}&{ .004}&{.041}&{.752} & {$\mathbf{74.94\pm0.44}$}&{\textless.001}&\multicolumn{1}{C{1.3}}{\textless.001}&\multicolumn{1}{C{1.3}}{\textless.001}&{.002}  \\
    \end{tabular}
% }
\end{table*}

\section{Discussion}
\label{sec:discussion}

\subsection{Quality of generated samples}

Thus far, we have presented a framework for generating realistic light curves that can be used to improve the classification of real astronomical objects. In the entire process, we constantly generate sets of samples that are then compared to the set of real samples, computing global metrics that indicate the quality of the model based on the distance between the sets. However, no metrics to evaluate the quality of individual samples have been mentioned.

In fact, Figure \ref{fig:GenSamples} shows that although the generated samples look generally realistic, there can be samples that present artifacts, making them not the best candidates for the classes they intend to represent. While these could be easily solved by applying truncation techniques on latent space of G, it would not be informative about the quality of the individual samples themselves, impeding us from learning what makes a sample look realistic.

The selected metric to evaluate individual sample quality is the \textit{realism score} \citep{impr2019}, computed over the manifold representation used for the $\mathbf{D}$ and $\mathbf{R}$ metrics. Given a generated feature sample $\phi_g$ and a set of real samples $\mathbf{\Phi}_r = \{\phi_r\}$, the similarity between $\phi_g$ and the real manifold $\Phi_r$ is calculated as: 

\begin{equation}
\label{eq:rscore}
    \mathcal{R}(\phi_g,\Phi_r) = \max_{\phi_r \in \mathbf{\Phi_r}} \Bigg\{ \displaystyle \frac{\left\Vert{NND_{k}(\phi_r)}  \right\Vert_{2}}{\left\Vert{\phi_r - \phi_g} \right\Vert_{2}} \Bigg\}
\end{equation}

\noindent 
where $NND_k(\phi)$ is the distance from $\phi$ to its $k$-th nearest neighbor within the corresponding manifold. Equation \ref{eq:rscore} compares the radii of the KNN induced hyperspheres with center in $\phi_r$  to the distance between $\phi_r$ and the sample $\phi_g$. Naturally, if $\phi_g$ does not belong to any of the hyperspheres, $\mathcal{R}$ will be low, and its value will increase the closer $\phi_g$ is to any $\phi_r$. 

\begin{figure*}[!htbp]
\centering
\includegraphics[width=\textwidth]{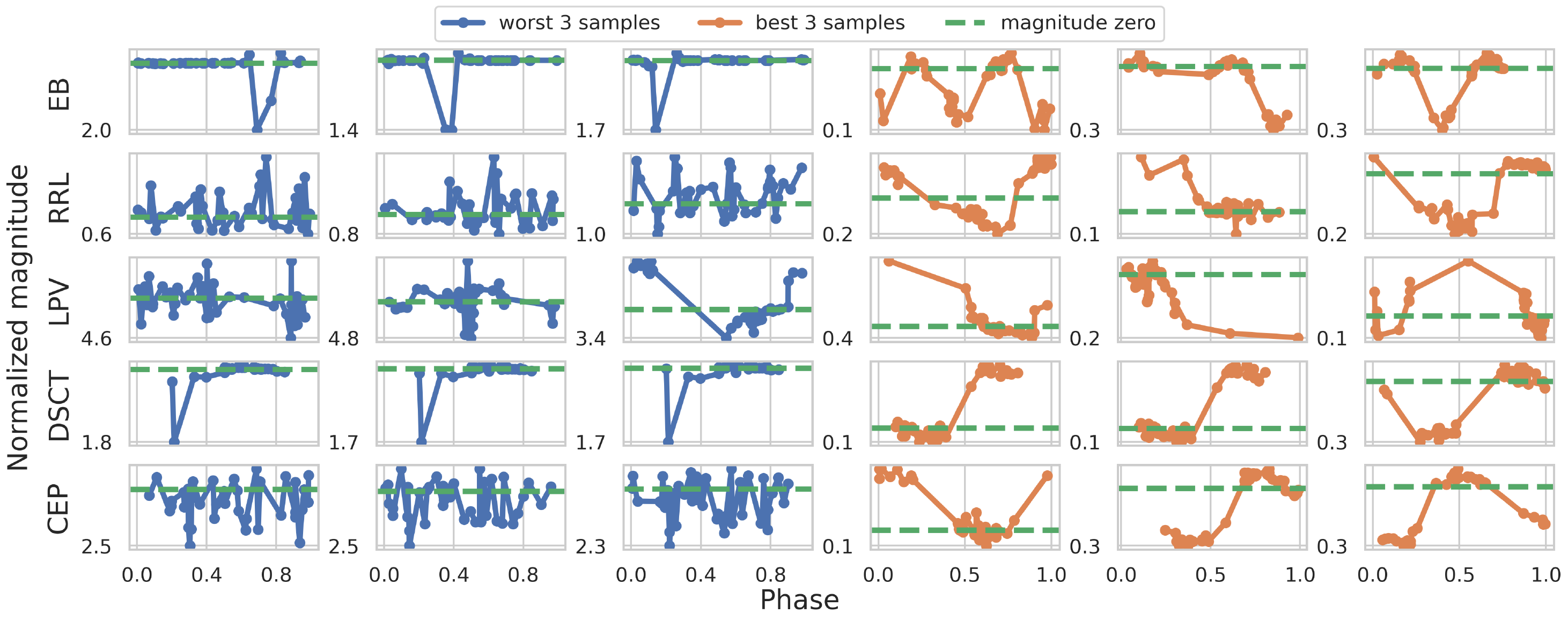}
\caption{Realism score ranking of the ZTF generated light curves. To rank the samples, we first generated a replication of the $D_{train}^u$, computed their realism score, and then selected the best and worst samples from the sorted realism scores. }
\label{fig:RealismZTF}
\end{figure*}

The effect of ranking the generated samples of the ZTF dataset by \textit{realism score} is shown in Figure \ref{fig:RealismZTF}.

Because it can successfully identify artifacts that could be filtered out of the dataset, we would in principle expect that using a \textit{realism score} filtering would improve our results even further. However, this is not the case. Empirically, we found no statistical differences when applying this filtering to our generated datasets. We hypothesize that these artifacts, although undesirable, are not crucial when defining the decision boundaries of the problem, hence, they have little impact on the classification accuracy. Moreover, strongly filtered datasets cause a drop in the classification accuracy, probably caused by their over-constrained diversity.

\subsection{Classification results}
\label{subsec:discclf}

% Why TRTR augmentations are useless
\paragraph{Soft-warping transformations}
Regarding the effects of the proposed soft-warping augmentations for classification, we can see that despite the fact that they create plausible light curves, they do not show improvements in the classification task. We hypothesize that the diversity added to the dataset by these transformations is not substantial enough for the classifiers to benefit from it.

% Why we think gamma is cool, and why we do not tune it.
\paragraph{$\gamma$-resampling}

The results suggest that the proposed resampling offers a clear improvement upon uniform resampling for GAN training. We believe that this improvement comes from the delay in the GAN overfitting, providing more potentially good models to choose from before the GAN completely overfits. %Under this hypothesis, we suspect that applying the resampling is equivalent to increasing the iteration frequency we use to validate the model, which allows capturing better models in anticipation of overfitting.
With respect to the no-resampling model, Figure \ref{fig:delta_tstr_vs_it} shows that models trained $\gamma=0$ and $\gamma=0.25$ reach comparable accuracies, consistently with the fact that the resampling block does not add any extra information. Using the resampling block can offer a more stable training that reaches similar performance in a shorter training time. This can be particularly relevant if the defined iteration horizon is not long enough to capture the peak accuracy as in figure \ref{fig:delta_tstr_vs_it}b. For this reason, we do not think that $\gamma$ should be tuned thoroughly, and we set it to $\gamma=0.25$, placing the $\mathcal{G}$-score peak within the extent of training iterations, earlier than the peak of $\gamma=0$ but later than that of $\gamma=1$.

% G-Scores vs val acc
\paragraph{$\mathcal{G}$-score}
For the model selection criterion, the correlation between the metrics and the classification results validate the $\mathcal{G}$-score as a metric to evaluate the quality of the generated samples. Using this metric instead of the validation accuracy, it is interesting because of the subtle improvements in TSTR. It also offers faster computation times: computing $\mathcal{G}$-score is approximately six times faster than computing the validation TSTR accuracy.

% Why we think G-score is cool
We hypothesize that these subtle improvements come from the robustness of the G-score against overfitting. While $\mathcal{G}$-score compares $\mathcal{D}_{gen}^u$ to the entire training set $\mathcal{D}_{train}$, the validation accuracy score is computed on the small dataset $\mathcal{D}_{val}$ for evident reasons. Hence, it is more susceptible to overfitting. A fact that reinforces this hypothesis is the consistently lower variance of the models selected with the $\mathcal{G}$-score criterion   compared to validation accuracy. 
% Drawbacks
On the other hand, computing the $\mathcal{G}$-score also has some drawbacks related to the normalization step restrictions. Since the normalization requires the minimum and maximum value of the \textbf{D} and \textbf{R} metrics, we cannot compute the $\mathcal{G}$-score during the training time, and we must first completely train the models. In addition to this, it only allows for comparison between different candidates of the same run, not permitting comparisons between different runs that likely have different normalization parameters.
%\pp{great point and well explained}
%\ggj{Thanksssssss}

\subsection{Alternative to $\mathcal{G}$-score}

Evaluating generative models by fidelity and diversity can be posed as a multi-objective problem. Thus, we provide an alternative to the $\mathcal{G}$-score that considers both objectives ($\mathbf{D}\&\mathbf{R}$) simultaneously, according to the problem's nature. 

As an alternative to evaluate all candidates with TSTR validation accuracy, we propose evaluating only candidates that lie on the \textit{Pareto frontier}\footnote{In multi-objective optimization, the Pareto frontier is the set of all the Pareto optimal solutions. A Pareto optimal solution is defined as a solution that cannot be improved in any individual objective without worsening others.} of the raw macro-density and macro-recall. For example, in the case of the Catalina dataset, doing so would imply evaluating approximately $1/4$ of total candidates.

\begin{figure}[!htbp]
\centering 
\includegraphics[width=.5\textwidth]{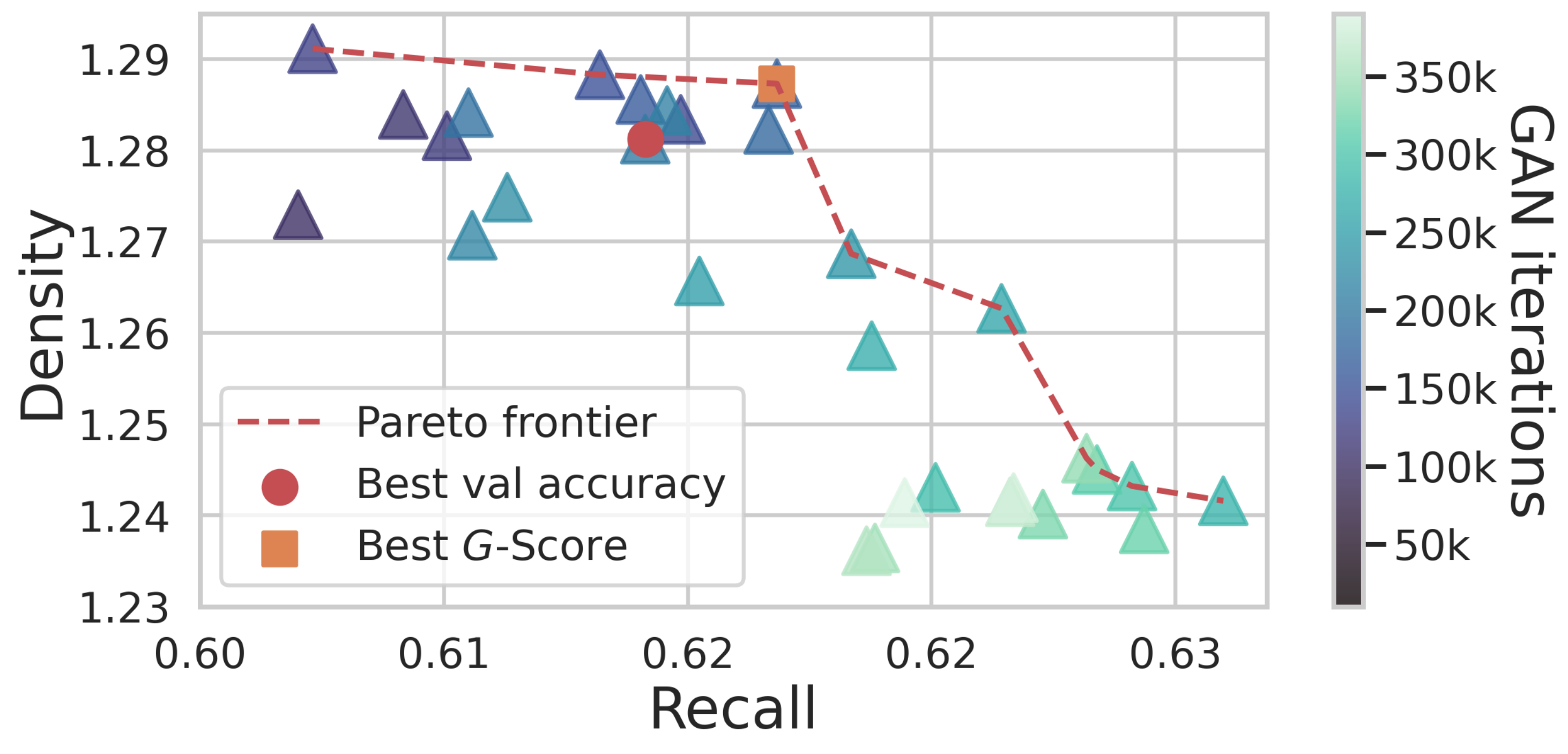}
\caption{Macro density and recall metrics for the Catalina dataset. Each point corresponds to the average of 5 independent computation of density and recall for a single GAN model.}
\label{fig:ParetoFrontierCatalina}
\end{figure}

The disposition of the optima for the Catalina dataset is shown in Figure \ref{fig:ParetoFrontierCatalina}. Interestingly, the model selected with the validation accuracy criterion is in the sub-optimal region which supports the idea of overfitting explained in Section \ref{subsec:discclf}. On the other hand, the model selected with $\mathcal{G}$-score belongs to the Pareto frontier, which is not necessarily guaranteed considering the extra normalization step included in the computation of the $\mathcal{G}$-score.

Using this alternative offers an attractive advantage. Not performing the normalization step of the $\mathcal{G}$-score allows for comparing different GAN setups in the $\mathbf{D}\mathbf{R}$ plane, which could also be used to perform hyperparameter optimization of the models. In this scenario, we first need to identify the models that lie in the Pareto frontier considering all the $\mathbf{D}\&\mathbf{R}$ scores and then evaluate these candidates based on the validation TSTR score to choose an operating point.

\section{Conclusions}
\label{sec:conclusions}

% Brief global summary 
In this work, we have presented a GAN-based data augmentation methodology for astronomical time-series, to improve the classification accuracy of periodic variable stars by mitigating the problems of small and imbalanced astronomical datasets.

% Detailed summary
Using our methodology, we can generate diverse synthetic datasets of irregularly sampled time-series that capture the original training sets' properties and leverage their diversity to outperform classifiers trained on real data. Motivated by the rapid overfitting of our generative model in this unbalanced setup, we propose a resampling technique ($\gamma$-resampling) to mitigate this behavior. Also, inspired by the incapability of FID to measure this overfitting, we propose a novel evaluation metric ($\mathcal{G}$-score) that correlates with TSTR classification accuracy; hence it helps select a generative model among the possible candidates saved during training. 

% Astronomical implications - Mention Vera Rubin, LC Classifier
The proposed model could be extended to work with classifiers that are currently operating in real-time such as the ALeRCE light curve classifier \citep{late2021}, boosting its performance on the ZTF stream and eventually on its successor, the Vera C. Rubin Observatory Legacy Survey of Space and Time (LSST; \citealt{lsst2019}), contributing to understanding the tridimensional structure and formation of our galaxy and its neighbors.

\subsection{Future Work}

Although effective in this simplified setup, the presented methodology could be improved by upgrading it to a scenario where the input data has a variable length. This upgrade should involve recent GAN models that include recurrent neural networks in their architectures, such as \citet{timeseriesgan2019} or \citet{sigwgan2020}. In addition, the generation of data with variable length should also be addressed.

% Talk about distribution of the conditional parameters
Regarding conditional generation, we used the class-conditional parameter to generate datasets with uniform class distributions. Although our model permits other conditional parameters such as amplitude, in all experiments we replicated the distribution of their real counterparts. An interesting extension of the work could include analyzing how the results vary depending on the generated conditional distribution of these parameters, and other physical parameters that may be relevant to include.

% Talk about sampling methods
Finally, all our synthetic datasets were generated by sampling $z$ from a multivariate Gaussian distribution related to data samples generation. Evaluating different sampling methods, such as those presented in \citet{impr2019}, and inspecting how they affect the qualitative and quantitative results, could be an exciting path to follow.

% Momantai

\section{Acknowledgments}

The authors acknowledge support from the National Agency of Research and Development’s Millennium Science Initiative through grant IC12009, awarded to the Millennium Institute of Astrophysics (GG, PE) and from the National Agency for Research and Development (ANID) grants: FONDECYT Regular $\#$1220829 (PE), and Magister Nacional/2019-22190949 (GG). This work was funded in part by the Institute for Applied Computational Science (IACS), Harvard University (PP).
% \appendix

\newpage
\bibliography{ref}{}
\bibliographystyle{aasjournal}


% \listofchanges

\end{document}